\documentclass{IEEEtran}
\usepackage[latin9]{inputenc}
\usepackage{mathrsfs}
\usepackage{amsmath}
\usepackage{amsthm}
\usepackage{graphicx}

\makeatletter
\theoremstyle{plain}
\newtheorem{thm}{\protect\theoremname}
\theoremstyle{plain}
\newtheorem{lem}[thm]{\protect\lemmaname}

\usepackage{amssymb}
\theoremstyle{definition}
\newtheorem{eg}{Example}

\theoremstyle{plain}
\newtheorem{mythm}{Theorem}

\theoremstyle{plain}
\newtheorem{mycor}{Corollary}

\theoremstyle{definition}
\newtheorem{remark}{Remark}

\theoremstyle{definition}
\newtheorem{mydef}{Definition}

\makeatother

\providecommand{\lemmaname}{Lemma}
\providecommand{\theoremname}{Theorem}

\begin{document}

\title{Matrix Exponential Learning Schemes with Low Informational Exchange}

\author{Wenjie~Li and Mohamad~Assaad\thanks{This paper was presented in part at IEEE 19th International Workshop
on Signal Processing Advances in Wireless Communications (SPAWC),
Kalamata, Greece, June 2018 \cite{li2018spawc}.}\thanks{W. Li and M. Assaad are with the Laboratoire des Signaux et Systèmes
(L2S, UMR CNRS 8506), CentraleSupélec, France (e-mail: wenjie.li@lss.centralesupelec.fr;
mohamad.assaad@centralesupelec.fr).}
\thanks{W. Li and M. Assaad are also with the TCL Chair on 5G, CentraleSupélec,
France.}
\thanks{This research has been performed in the framework of the Horizon 2020
project ONE5G (ICT-760809) receiving funds from the European Union.
The authors therefore would like to acknowledge the contributions of their
colleagues in the project, although the views and work expressed in this
contribution are those of the authors and do not represent the project.}}
\maketitle
\begin{abstract}
We consider a distributed resource allocation problem in networks
where each transmitter-receiver pair aims at maximizing its local
utility function by adjusting its action matrix, which belongs to
a given feasible set. This problem has been addressed recently by
applying a matrix exponential learning (MXL) algorithm which has a
very appealing convergence rate. In this learning algorithm, however,
each transmitter must know an estimate of the gradient matrix of the
local utility. The knowledge of the gradient matrix at the transmitters
incurs a high signaling overhead especially that the matrix size increases
with the dimension of the action matrix. In this paper, we therefore
investigate two strategies in order to decrease the informational
exchange per iteration of the algorithm. In the first strategy, each
receiver sends at each iteration part of the elements of the gradient
matrix with respect to a certain probability. In the second strategy,
each receiver feeds back ``sporadically'' the whole gradient matrix.
We focus on the analysis of the convergence of the MXL algorithm to
optimum under these two strategies. We prove that the algorithm can
still converge to optimum almost surely. Upper bounds of the average
convergence rate are also derived in both situations with general
step-size setting, from which we can clearly see the impact of the
incompleteness of the feedback information. The proposed algorithms
are applied to solve the energy efficiency maximization problem in
a multicarrier multi-user MIMO network. Simulation results further
corroborate our claim.
\end{abstract}

\section{Introduction}

\label{sec:intro}

Recently, there has been a growing interest in adopting cooperative
and non-cooperative game theoretic approaches to model many communications
and networking problems, such as power control and resource sharing
in wireless/wired and peer-to-peer networks, cognitive radio systems,
and distributed routing, flow, and congestion control in communication
networks, see, for example \cite{bennis2013self,lasaulce2011game,12,Scutari2009MIMOWF,ji2007Cognitive}.

This paper deals with a resource allocation problem in networks where
each user tries to maximize its local utility. More specifically,
we consider the multi-user, multi-carrier MIMO networks, each user
controls its signal covariance matrix and the local utility function
is defined as the energy efficiency of each user \cite{9}. This problem
has been addressed very recently in \cite{1} by applying the matrix
exponential learning (MXL) technique, of which the convergence to
Nash Equilibrium (NE) has been well demonstrated. The MXL-based algorithm
is attractive because of its fast convergence to NE. As most of the
gradient based methods \cite{snyman2005practical,Bertsekas2010},
the gradient of the utility function should be estimated by receivers
and then sent back to transmitters as signaling information. Another
possibility is to let the transmitters compute the gradient, which
requires a full CSI knowledge of all direct and cross links and hence
requires more signaling overhead than feeding back the gradient as
we will see later in the paper. Although feeding back the gradient
decreases the signaling overhead, it is still of huge burden of the
network. In multi-carrier MIMO networks, the feedback is a gradient
matrix with its size related to the number of transmission antennas
and the number of carriers. When many users are present in the network,
such signaling overhead may introduce a huge traffic burden.

For the above reason, our main goal is to investigate some modified
MXL-based algorithm which requires less amount of signaling overhead
and ensures the convergence to NE. Two possible variants are considered:
\emph{i}) each receiver feedback at each iteration only \emph{part}
of the elements of the gradient matrix instead of the full gradient
matrix, the elements of the action matrix do not update if the associated
element of the gradient matrix is not available; \emph{ii}) each receiver
sporadically feedback the whole gradient matrix, instead of feedbacking
at each iteration, thus not all of the transmitters are able to update
their action at the same time.

As an extension of the work in \cite{1}, we have analyzed the two
variants of the MXL algorithms and shown that they converge to NE
almost surely. In both settings, we also derive the evolution of the
average quantum Kullback-Leibler divergence \cite{vedral2002role},
of which the upper bounds are obtained. The theoretical results provide
a quantitative description of the impact of incompleteness on the
convergence rate. Our analysis is much more complicated compared with
that in \cite{1}, as we have introduced some additional random terms
into the algorithms, \emph{i.e.}, random incompleteness of the feedback
information and sporadic update of the action. As a consequence, we
have to define some additional stochastic noise to be analyzed by
applying Doob's martingale inequality. The sporadic algorithm is more
challenging to be analyzed by the fact that the update of action takes
place at random time slot for each transmitter, and the step-size
is also generated in a random manner. Some concentration inequality,
such as Chernoff bound, has to be used to show the convergence of
the sporadic algorithm. Another contribution of this paper is that
we consider general forms of the step-size $\gamma_{n}$ to derive
the convergence rate, whereas $\gamma_{n}=\alpha/n$ is considered
in \cite{1}. Simulations further justify our results.

The rest of the paper is organized as follows. Section~\ref{sec:MXL}
presents the system model and briefly describes the MXL algorithm
proposed in \cite{1}. Section~\ref{sec:MXL-with-Incomplete} presents
the MXL algorithm with incomplete feedback information and presents
its convergence. The sporadic MXL is analyzed in Section~\ref{sec:Asynchronous-MXL}.
Section~\ref{sec:Numerical-Example} provides some numerical examples
and Section~\ref{sec:Conclusion} concludes this paper.

Throughout this paper, we denote $\mathbf{X}=\textrm{diag}\left(\mathbf{X}_{k}\right)_{k=1}^{K}$
and $\mathbf{X}_{-k}=\left(\mathbf{X}_{j}\right)_{j\neq k}$. The
$m$-th eigenvalue of a matrix $\mathbf{X}$ is denoted by $\textrm{eig}_{m}\left(\mathbf{X}\right)$,
$\Vert\mathbf{X}\Vert=\textrm{tr}\left(\mathbf{X}\right)=\sum_{m=1}^{M}\left|\textrm{eig}_{m}\left(\mathbf{X}\right)\right|$
denotes the trace norm of $\mathbf{X}$, and $\Vert\mathbf{X}\Vert_{\infty}=\max_{m}\left|\textrm{eig}_{m}\left(\mathbf{X}\right)\right|$
represents the singular norm.

\section{System Model and MXL Algorithm}

\label{sec:MXL} 

This section presents the basic system model and briefly recalls the
MXL algorithm proposed in \cite{1}.

\subsection{Problem formulation}

This section introduces the general mathematical system model.

Consider a finite set of transmitter-receiver links $\mathcal{K}=\left\{ 1,\ldots,K\right\} $.
Each transmitter $k$ needs to properly control its action matrix
$\mathbf{X}_{k}$ in order to maximize its local utility $u_{k}\left(\mathbf{X}_{1},...,\mathbf{X}_{K}\right)$,
which depends on the action of all the transmitters. 

Notice that the instant local utility $\widetilde{u}_{k}$ may be
affected by some stochastic environment state $\mathbf{S}$, \emph{e.g}.,
time-varying channels. In such situation, we consider $u_{k}\left(\mathbf{X}_{1},...,\mathbf{X}_{K}\right)=\mathbb{E}_{\mathbf{S}}\left[\widetilde{u}_{k}\left(\mathbf{X}_{1},...,\mathbf{X}_{K};\mathbf{S}\right)\right]$.
For any link $k$, its local utility $u_{k}$ is assumed to be \emph{concave}
and \emph{smooth} in $\mathbf{X}_{k}$. 

Consider the same setting as in \cite{1}, the action matrix $\mathbf{X}_{k}$
has to belong to a pre-defined feasible action set $\mathscr{X}_{k}$,
\emph{i.e.},
\[
\mathscr{X}_{k}=\left\{ \mathbf{X}_{k}\succcurlyeq0:\Vert\mathbf{X}_{k}\Vert\leq A_{k}\right\} ,
\]
with $A_{k}$ is a positive constant. 

Our problem is then, $\forall k\in\mathcal{K}$,
\begin{equation}
\begin{cases}
\textrm{maximize} & u_{k}(\mathbf{X}_{1},...,\mathbf{X}_{K}),\\
\textrm{subject to} & \mathbf{X}_{k}\in\mathscr{X}_{k}.
\end{cases}\label{eq:game}
\end{equation}
The solution of the problem \eqref{eq:game} is the well known Nash
equilibrium (NE) $\mathbf{X}^{*}=\left(\mathbf{X}_{1}^{*},\ldots,\mathbf{X}_{K}^{*}\right)\in\mathcal{X}$,
which satisfies 
\[
u_{k}\left(\mathbf{X}^{*}\right)\geq u_{k}\left(\mathbf{X}_{k},\mathbf{X}_{-k}^{*}\right),\:\forall\mathbf{X}_{k}\in\mathcal{X}_{k},\forall k\in\mathcal{K}.
\]
Note that the existence of NE can be guaranteed as $u_{k}$ is concave
with respect to $\mathbf{X}_{k}$ \cite{debreu1952social}. We will
show in Section \ref{subsec:Multicarrier,-multi-user-MIMO} an example
where \eqref{eq:game} admits a NE.

\subsection{MXL algorithm }

In order to solve the above problem, a solution has been proposed
recently in \cite{1}, named Matrix Exponential Learning (MXL). This
MXL algorithm is interesting thanks to its robustness against the
stochastic environment and its fast convergence to NE. This section
briefly describes this algorithm.

Suppose that each transmitter $k$ is able to get a noisy estimation
$\widehat{\mathbf{V}}_{k}\left(n\right)$ of the individual gradient
$\mathit{\mathbf{V}_{k}\left(n\right)}=\partial u_{k}/\partial\mathbf{X}_{k}\left(n\right)$.
According to the MXL algorithm \cite{1}, each user $k$ updates its
action at each iteration $n\geq1$
\begin{align}
\mathbf{X}_{k}\left(n\right) & =A_{k}\frac{\exp\left(\mathbf{Y}_{k}\left(n-1\right)\right)}{1+\Vert\exp\left(\mathbf{Y}_{k}\left(n-1\right)\right)\Vert},\label{eq:mxl}\\
\mathbf{Y}_{k}\left(n\right) & =\mathbf{Y}_{k}\left(n-1\right)+\gamma_{n}\widehat{\mathbf{V}}_{k}\left(n\right),\label{eq:gradient_descent}
\end{align}
in which $\gamma_{n}$ is a pre-defined vanishing step-size and $\mathbf{Y}_{k}\left(n\right)$
is an intermediate matrix variable with $\mathbf{Y}_{k}\left(0\right)$
an arbitrary Hermitian value. Notice that \eqref{eq:gradient_descent}
can be seen as a step of the gradient ascent method and \eqref{eq:mxl}
ensures that $\mathbf{X}_{k}\left(n\right)$ meets the action constraint.
Interested readers may refer to \cite{1} for the detailed properties
of the step \eqref{eq:mxl}, here $\mathbf{Y}_{k}\left(n\right)$
can be seen as the auxiliary dual variable of the primal variable
$\mathbf{X}_{k}\left(n\right)$. Note that the dimension of the matrices
$\mathbf{X}_{k}\left(n\right)$, $\mathbf{Y}_{k}\left(n\right)$,
and $\widehat{\mathbf{V}}_{k}\left(n\right)$ is the same.

To guarantee the performance of the MXL algorithm, the following assumptions
are made, which are common in the typical stochastic approximation
problems \cite{borkar2008stochastic,kushner2012stochastic} :

\emph{A1}. $\widehat{\mathbf{V}}\left(n\right)=\mathbf{V}\left(n\right)+\mathbf{Z}\left(n\right)$
where $\mathbf{Z}\left(n\right)$ is an additive stochastic noise.
Introduce $\mathcal{F}_{n}=\left\{ \mathbf{Z}\left(0\right),\ldots,\mathbf{Z}\left(n\right)\right\} $,
then any element $Z_{i,j}\left(n\right)$ satisfies $\mathbb{E}\left[Z_{i,j}\left(n\right)\mid\mathcal{F}_{n-1}\right]=\mathbf{0}$
and $\mathbb{E}\left[Z_{i,j}^{2}\left(n\right)\mid\mathcal{F}_{n-1}\right]\leq\sigma^{2}$.

\emph{A2}. The positive vanishing step-size $\gamma_{n}$ satisfies
$\sum_{n=1}^{\infty}\gamma_{n}=\infty$ and $\sum_{n=1}^{\infty}\gamma_{n}^{2}<\infty$.

\subsection{Multicarrier, multi-user MIMO system\label{subsec:Multicarrier,-multi-user-MIMO}}

This section provides an example in which the assumptions on the concavity
of the local utility functions are satisfied and NE exists.

We consider the application of the MXL algorithm in a multicarrier,
multi-user MIMO network with $K$ transmitter-receiver links. During
the communication, $S$ orthogonal subcarriers are used, \emph{e.g.},
in OFDM systems. Each transmitter is equipped with $N_{t}$ transmission
antennas and each receiver has $N_{r}$ reception antennas. For any
$k\in\mathcal{K}$, $j\in\mathcal{K}$, and $s\in\mathcal{S}$, let
the matrix $\mathbf{H}_{kjs}\in\mathbb{C}^{N_{r}\times N_{t}}$ describe
the channel during the communication between transmitter $k$ any
receiver $j$ over subcarrier $s$. The channel matrix over all subcarriers
is then $\mathbf{H}_{kj}=\textrm{diag}\left(\mathbf{H}_{kjs}\right)_{s=1}^{S}$
of size $N_{r}S\times N_{t}S$. In this paper, we assume that $\mathbf{H}_{kj}$
is stochastic, time-varying, ergodic and Gaussian. 

Each transmitter $k$ controls its signal covariance matrix $\mathbf{Q}_{k}$
in order to maximize its own energy efficiency (EE) defined as $\textrm{EE}_{k}=r_{k}/\left(\textrm{tr}\left(\mathbf{Q}_{k}\right)+P_{\textrm{c}}\right)$,
where \emph{i}) $r_{k}$ denotes the Shannon-achievable rate, \emph{i.e}.,
$r_{k}=\log\det\left(\mathbf{I}+\sum_{j\in\mathcal{K}}\mathbf{H}_{jk}\mathbf{Q}_{k}\mathbf{H}_{jk}^{\dagger}\right)-\log\det\left(\mathbf{I}+\sum_{j\in\mathcal{K}\setminus\left\{ k\right\} }\mathbf{H}_{jk}\mathbf{Q}_{k}\mathbf{H}_{jk}^{\dagger}\right)$;
\emph{ii}) $\textrm{tr}\left(\mathbf{Q}_{k}\right)+P_{\textrm{c}}$
is the total power consumed by transmitter $k$, including the transmission
power $\textrm{tr}\left(\mathbf{Q}_{k}\right)$ and the total circuit
consumption power $P_{\textrm{c}}$ \cite{bjornson2015optimal}. Note
that $\mathbf{Q}_{k}=\textrm{diag}\left(\mathbf{Q}_{ks}\right)_{s=1}^{S}$
with $\mathbf{Q}_{ks}\in\mathbb{H}^{N_{t}\times N_{t}}$ the covariance
matrix over subcarrier $s$, we use $\mathbb{H}$ to denote the set
of Hermittian matrices. Since the covariance matrix is positive semi-definite,
we have $\mathbf{Q}_{ks}\succcurlyeq0$. Introduce $P_{\max}$ the
maximum transmission power of any transmitter $k$, then we have $\textrm{tr}\left(\mathbf{Q}_{k}\right)\leq P_{\max}$.
Thus $\mathbf{Q}_{k}\in\mathscr{X}_{k}$ with $A_{k}=P_{\max}$, $\forall k\in\mathcal{K}$.
Recall that our aim is to make each transmitter $k$ maximize its
$\textrm{EE}_{k}$, which depends on the channel state of the network
as well as the signal covariance matrices of all the transmitters.
Hence, in this situation, our problem \eqref{eq:game} turns to, $\forall k\in\mathcal{K}$,
\begin{equation}
\begin{cases}
\textrm{maximize} & \mathbb{E}\left[\textrm{EE}_{k}\left(\mathbf{Q}_{1},...,\mathbf{Q}_{K};\mathbf{H}\right)\right],\\
\textrm{subject to} & \mathbf{Q}_{k}\in\left\{ \textrm{diag}\left(\mathbf{Q}_{ks}\right)_{s=1}^{S}:\mathbf{Q}_{ks}\in\mathbb{H}^{N_{t}\times N_{t}},\right.\\
 & \qquad\left.\mathbf{Q}_{ks}\succcurlyeq0,\textrm{ and }\sum_{s=1}^{S}\textrm{tr}\left(\mathbf{Q}_{ks}\right)\leq P_{\max}\right\} .
\end{cases}\label{eq:game-1}
\end{equation}
At a first look, $\textrm{EE}_{k}\left(\mathbf{Q}_{1},...,\mathbf{Q}_{K};\mathbf{H}\right)$
is not a concave function of $\mathbf{Q}_{k}$. However, EE can be
concave after a variable change: as presented in \cite{9,1}, we apply
the following adjusted action matrix $\mathbf{X}_{k}$ instead of
$\mathbf{Q}_{k}$, such that $\widetilde{u}_{k}$ is concave with
respect to $\mathbf{X}_{k}$,\emph{ i.e}.,
\[
\mathbf{X}_{k}=\frac{P_{\textrm{c}}+P_{\textrm{max}}}{P_{\textrm{max}}}\frac{\mathbf{Q}_{k}}{P_{\textrm{c}}+\textrm{tr}(\mathbf{Q}_{k})}.
\]
Using the transformation, we have \cite{9}
\begin{align*}
\widetilde{u}_{k}(\mathbf{X}_{1},...,\mathbf{X}_{K};\mathbf{H}) & =\frac{P_{\textrm{c}}+P_{\textrm{max}}\left(1-\textrm{tr}(\mathbf{X}_{k})\right)}{P_{\textrm{c}}\left(P_{\textrm{c}}+P_{\textrm{max}}\right)}\\
 & \:\cdot\log\det\left(\mathbf{I}+\frac{P_{\textrm{c}}P_{\textrm{max}}\mathbf{\widetilde{H}}_{k}X_{k}\mathbf{\widetilde{H}}_{k}^{\dagger}}{P_{\textrm{c}}+P_{\textrm{max}}\left(1-\textrm{tr}(\mathbf{X}_{k})\right)}\right),
\end{align*}
where $\mathbf{\widetilde{H}}_{k}=\left(\mathbf{I}+\sum_{j\neq k}\mathbf{H}_{jk}\mathbf{Q}_{k}\mathbf{H}_{jk}^{\dagger}\right)^{-\frac{1}{2}}\mathbf{H}_{kk}$
is the effective channel matrix. By definition, we have $\mathbf{X}_{k}\in\mathbb{H}^{N_{t}S\times N_{t}S}$
and $\mathbf{X}_{k}\succcurlyeq0$ . As $\textrm{tr}\left(\mathbf{Q}_{k}\right)\leq P_{\max}$,
we can deduce that $\textrm{tr}\left(\mathbf{X}_{k}\right)\leq1$.
Thus $\mathbf{X}_{k}\in\mathscr{X}_{k}$ with $A_{k}=1$, $\forall k\in\mathcal{K}$.
More precisely, our problem \eqref{eq:game-1} turns to, $\forall k\in\mathcal{K}$,
\begin{equation}
\begin{cases}
\textrm{maximize} & \mathbb{E}\left[\widetilde{u}_{k}(\mathbf{X}_{1},...,\mathbf{X}_{K};\mathbf{H})\right],\\
\textrm{subject to} & \mathbf{X}_{k}\in\left\{ \textrm{diag}\left(\mathbf{X}_{ks}\right)_{s=1}^{S}:\mathbf{X}_{ks}\in\mathbb{H}^{N_{t}\times N_{t}},\right.\\
 & \qquad\left.\mathbf{X}_{ks}\succcurlyeq0,\textrm{ and }\sum_{s=1}^{S}\textrm{tr}\left(\mathbf{X}_{ks}\right)\leq1\right\} ,
\end{cases}\label{eq:game-1-1}
\end{equation}
which admits a NE, as the concavity assumption is satisfied. The MXL
algorithm can be thus easily applied in this system. 

\begin{remark}As pointed out in \cite{1}, a first implementation
issue is that $\mathbf{X}_{k}\left(n\right)$ should keep Hermitian
to satisfy the feasibility constraint. By the basic steps of the MXL
algorithm, we find that the estimation of the gradient matrix $\widehat{\mathbf{V}}_{k}\left(n\right)$
has to be Hermitian, which cannot be true due to the additive noise
$\mathbf{Z}\left(n\right)$. A simple solution we propose here is
then using $\left(\widehat{\mathbf{V}}_{k}\left(n\right)+\widehat{\mathbf{V}}_{k}^{\dagger}\left(n\right)\right)/2$
instead of $\widehat{\mathbf{V}}_{k}\left(n\right)$. Another issue
is the signaling overhead introduced by the MXL algorithm, we highlight
this problem in the next section.\end{remark}

\subsection{Signaling overhead}

In the MXL algorithm, each transmitter $k$ needs to have the full
knowledge of the gradient matrix $\widehat{\mathbf{V}}_{k}\left(n\right)$.
There are two possible options to obtain $\widehat{\mathbf{V}}_{k}\left(n\right)$:

The first option is to make each transmitter compute the gradient.
For brevity, we skip the derivation of the gradient which is straightforward
\cite{9}. For each transmitter $k$, the computation of $\partial\widetilde{u}_{k}/\partial\mathbf{X}_{k}\left(n\right)$
requires the knowledge of the channel matrices $\mathbf{H}_{jk}$
with all $j\in\mathcal{K}$, including the direct link and all the
cross links to receiver $k$. The amount of necessary information
is then $KSN_{t}N_{r}$.

The second option is to make each receiver $k$ compute the gradient
and then directly feedback the gradient matrix $\widehat{\mathbf{V}}_{k}\left(n\right)$.
Note that the dimension of $\widehat{\mathbf{V}}_{k}\left(n\right)$
is the same as $\mathbf{X}_{k}\left(n\right)$, thus the signaling
overhead in this situation is of size $SN_{t}^{2}$. 

It is obvious that the second option requires less signaling overhead
as the number of links is high. For this reason, we focus on the second
option in this paper. It is notable that feeding back $SN_{t}^{2}$
elements is still a huge burden of the network. This is the reason
why we focus on the variant of the MXL algorithm with less signaling
overhead.

\section{MXL with Incomplete Feedback}

\label{sec:MXL-with-Incomplete}

\label{subsec:Main-result_I}

In this section, we consider a first strategy to reduce the signaling
overhead: each receiver $k$ does not send every element of the matrix
$\widehat{\mathbf{V}}_{k}\left(n\right)$. More precisely:

\emph{S1}. Each element $\widehat{V}_{k,i,j}\left(n\right)$ of $\widehat{\mathbf{V}}_{k}\left(n\right)$
is sent with a constant probability $p_{\textrm{I}}\in\left(0,1\right)$
with $i\geq j$ and $\widehat{V}_{k,j,i}\left(n\right)=\widehat{V}_{k,i,j}\left(n\right)$.
The non-received elements are replaced by 0.

Notice that the transmitted elements of $\widehat{\mathbf{V}}_{k}\left(n\right)$
has symmetric positions in order to ensure that the received gradient
matrix is Hermitian. 

Introduce a symmetric matrix $\boldsymbol{\Delta}_{k}\left(n\right)=\left[\delta_{k,i,j}\left(n\right)\right]$
of the same dimension as $\widehat{\mathbf{V}}_{k}\left(n\right)$.
The elements of $\delta_{k,i,j}\left(n\right)$ are i.i.d. Bernoulli
random variables with $\mathbb{P}\left[\delta_{k,i,j}\left(n\right)=1\right]=p_{\textrm{I}}$
as $i\geq j$, \emph{i.e.}, $\delta_{k,i,j}\left(n\right)\sim\mathcal{B}\left(1,p_{\textrm{I}}\right)$.
Mathematically, the actually transmitted gradient matrix $\widehat{\mathbf{V}}_{k}^{\left(\textrm{I}\right)}\left(n\right)$
can be seen as the Hadamard (element-wise) product of matrices $\widehat{\mathbf{V}}_{k}\left(n\right)$
and $\boldsymbol{\Delta}_{k}\left(n\right)$, \emph{i.e.}, 
\[
\widehat{\mathbf{V}}_{k}^{\left(\textrm{I}\right)}\left(n\right)=\boldsymbol{\Delta}_{k}\left(n\right)\circ\widehat{\mathbf{V}}_{k}\left(n\right).
\]
In this situation, the MXL algorithm presented in Section \ref{sec:MXL}
can still be performed, with \eqref{eq:mxl}-\eqref{eq:gradient_descent}
replaced by 
\begin{align}
\mathbf{X}_{k}^{\left(\textrm{I}\right)}\left(n\right) & =A_{k}\frac{\exp\left(\mathbf{Y}_{k}^{\left(\textrm{I}\right)}\left(n-1\right)\right)}{1+\Vert\exp\left(\mathbf{Y}_{k}^{\left(\textrm{I}\right)}\left(n-1\right)\right)\Vert},\label{eq:mxl_i}\\
\mathbf{Y}_{k}^{\left(\textrm{I}\right)}\left(n\right) & =\mathbf{Y}_{k}^{\left(\textrm{I}\right)}\left(n-1\right)+\gamma_{n}\boldsymbol{\Delta}_{k}\left(n\right)\circ\widehat{\mathbf{V}}_{k}\left(n\right).\label{eq:gradient_i}
\end{align}

In the rest of this section, we investigate the almost surely convergence
of MXL-I (the MXL performed with incomplete feedback), as well as
the convergence rate of MXL-I. Note that we only present our analysis
with $A_{k}=1$ to lighten the equations. The general case can be
analyzed in the same way with rescaling. 

\subsection{\label{subsec:Convergence-of-MXL-I}Convergence of MXL-I}

The analysis in this section is performed under the following assumption:

\emph{A3}. $\mathbf{X}^{*}\in\mathcal{X}$ is globally stable, which
means that 
\begin{equation}
\textrm{tr}\left(\left(\mathbf{X}-\mathbf{X}^{*}\right)\mathbf{V}\left(\mathbf{X}\right)\right)\leq0,\:\forall\mathit{\mathbf{X}}\in\mathcal{X}.\label{eq:g_stable}
\end{equation}
Notice that the global stability implies the uniqueness of NE $\mathbf{X}^{*}$.

We consider the generalized quantum Kullback-Leibler divergence \cite{vedral2002role}
\begin{align}
\mathit{\mathsf{d}_{\textrm{KL}}\left(\mathbf{X}^{*},\mathbf{X}\right)} & =\textrm{tr}\left(\mathbf{X}^{*}\left(\log\mathbf{X}^{*}-\log\mathbf{X}\right)\right)\nonumber \\
 & \qquad+\left(1-\textrm{tr}\left(\mathbf{X}^{*}\right)\right)\log\frac{1-\textrm{tr}\left(\mathbf{X}^{*}\right)}{1-\textrm{tr}\left(\mathbf{X}\right)}\label{eq:d_kl-1}
\end{align}
to have a measure of the distance between the NE $\mathbf{X}^{*}$
and an arbitrary action $\mathbf{X}\in\mathcal{X}$. Note that $\mathit{\mathsf{d}_{\textrm{KL}}\left(\mathbf{X}^{*},\mathbf{X}\right)}$
is the Bregman divergence \cite{shalev2012online} associated with
the strictly convex generating function $h\left(\mathbf{X}\right)$,
with
\begin{align}
h\left(\mathbf{X}\right) & =\textrm{tr}\left(\mathbf{X}\log\mathbf{X}\right)+\left(1-\textrm{tr}\left(\mathbf{X}\right)\right)\left(\log1-\textrm{tr}\left(\mathbf{X}\right)\right),\label{eq:hx}
\end{align}
\begin{equation}
\mathsf{d}_{\textrm{KL}}\left(\mathbf{X}^{*},\mathbf{X}\right)=h\left(\mathbf{X}^{*}\right)-h\left(\mathbf{X}\right)-\left\langle \nabla h\left(\mathbf{X}\right),\mathbf{X}^{*}-\mathbf{X}\right\rangle .
\end{equation}
In fact, $h\left(\mathbf{X}\right)$ is a modified von Neumann entropy
\cite{vedral2002role}. According to the general property of Bregman
divergence, we know that $\mathsf{d}_{\textrm{KL}}\geq0$ is a measure
of difference, increasing with the difference between $\mathbf{X}^{*}$
and $\mathbf{X}$. Besides, $\mathsf{d}_{\textrm{KL}}=0$ if and only
if $\mathbf{X}^{*}=\mathbf{X}$. 

We are interested in the evolution of the divergence $\mathsf{d}_{\textrm{KL}}\left(\mathbf{X}^{*},\mathbf{X}^{\left(\textrm{I}\right)}\left(n\right)\right)$
to analyze the convergence of algorithm. To begin with our analysis,
we recall a preliminary property of the MXL algorithm that has been
presented in \cite{1}.
\begin{lem}
\label{lem:MXL_pre}In the MXL algorithm, we have 
\begin{align}
 & \mathsf{d}_{\textrm{KL}}\left(\mathbf{X}^{*},\mathbf{X}\left(n+1\right)\right)\leq\mathsf{d}_{\textrm{KL}}\left(\mathbf{X}^{*},\mathbf{X}\left(n\right)\right)\nonumber \\
 & \qquad+\gamma_{n}\mathrm{tr}\left(\left(\mathbf{X}\left(n\right)-\mathbf{X}^{*}\right)\widehat{\mathbf{V}}\left(\mathbf{X}\right)\right)+\gamma_{n}^{2}\left\Vert \widehat{\mathbf{V}}\left(\mathbf{X}\right)\right\Vert _{\infty}^{2}.
\end{align}
\end{lem}
\begin{IEEEproof}
See Appendix~\ref{subsec:Proof-pre}.
\end{IEEEproof}
Let us now establish our first theoretical result in this paper, which
is the almost sure convergence of our modified MXL algorithm. 

\begin{mythm}\label{thm:mxli_conv}Suppose that Assumptions A1-A3
are satisfied, then MXL-I converges to NE almost surely, i.e., $\mathit{d_{n}^{\left(\mathrm{I}\right)}}\rightarrow0$
a.s..\end{mythm}
\begin{IEEEproof}
See Appendix~\ref{subsec:Proof-convmxli}.
\end{IEEEproof}
Due to the presence of the additional stochastic term $\boldsymbol{\Delta}_{k}\left(n\right)$,
the analysis of MXL-I is more complicated than that of MXL presented
in \cite{1}. As we can see in the proof in Appendix~\ref{subsec:Proof-convmxli},
there is an additional term of stochastic noise that has to be analyzed
and cannot be done by applying the tool used in \cite{1}. In fact,
we develop a new analysis by using Doob's martingale inequality to
prove that such novel noise has the required similar property compared
with the classical noise term $\mathbf{Z}=\widehat{\mathbf{V}}\left(\mathbf{X}\right)-\mathbf{V}\left(\mathbf{X}\right)$.
The demonstration detail is provided in Appendix~\ref{subsec:Proof_sto_e1}.

\subsection{Convergence rate of MXL-I}

In this section, we are interested in the evolution of the expected
value of the divergence over all the stochastic items, \emph{i.e.},
$D_{n}^{\left(\mathrm{I}\right)}=\mathbb{E}\left[d_{n}^{\left(\mathrm{I}\right)}\right]=\mathbb{E}\left[\mathsf{d}_{\textrm{KL}}\left(\mathbf{X}^{*},\mathbf{X}^{\left(\mathrm{I}\right)}\left(n\right)\right)\right]$.
To simplify the analysis, we further assume that the NE $\mathbf{X}^{*}$
is strongly stable 

\emph{A4}. Given a constant $B>0$, the NE $\mathbf{X}^{*}$ is $B-$strongly
stable, \emph{i.e.}, 
\begin{equation}
\textrm{tr}\left(\left(\mathbf{X}-\mathbf{X}^{*}\right)\mathbf{V}\left(\mathbf{X}\right)\right)\leq-B\mathit{\mathsf{d}_{\textrm{KL}}\left(\mathbf{X}^{*},\mathbf{X}\right),\:\forall\mathit{\mathbf{X}}\in\mathcal{X}.}\label{eq:strong_stable}
\end{equation}

We aim to derive an upper bound of $D_{n}^{\left(\mathrm{I}\right)}$
in order to find the convergence rate of MXL-I. We can also observe
the influence of the incompleteness of the feedback information on
the convergence rate. 

Despite of the fact that we have an additional stochastic term $\boldsymbol{\Delta}\left(n\right)$,
it is also worth mentioning that we consider a general setting of
the step-size, while $\gamma_{n}=n^{-1}$ in \cite{1}. Our result
is presented in what follows.

\begin{mythm}\label{prop:rate_incomp}Assume that the assumptions
A1, A2, and A4 hold, the MXL-I algorithm is performed such that each
element of the gradient matrix is sent with a constant probability
$p_{\mathrm{I}}$. Define
\[
\varepsilon=\max\frac{\gamma_{n}-\gamma_{n+1}}{\gamma_{n}^{2}},
\]
then if 
\begin{equation}
\varepsilon<p_{\mathrm{I}}B<\frac{1}{\gamma_{1}},\label{eq:condi_i}
\end{equation}
we have 
\begin{equation}
D_{n}^{\left(\mathrm{I}\right)}\leq\lambda\gamma_{n},\textrm{ }\mathrm{with}\textrm{ }\lambda=\max\left\{ \frac{D_{1}^{\left(\mathrm{I}\right)}}{\gamma_{1}},\frac{p_{\mathrm{I}}C}{p_{\mathrm{I}}B-\varepsilon}\right\} .\label{eq:dni}
\end{equation}
\end{mythm}
\begin{IEEEproof}
See Appendix \ref{subsec:Proof-rate-i}. 
\end{IEEEproof}
Thanks to the fact that $\gamma_{n}$ is vanishing and the constant
$\lambda$ is bounded, we find that the upper bound of $D_{n}^{\left(\mathrm{I}\right)}$
is vanishing by Theorem~\ref{prop:rate_incomp}. We may consider
another definition of convergence.

\begin{mydef}\label{def:conv=0000E8mean}We say that the MXL algorithm
converges \emph{in mean square} to NE if $\mathbb{E}\left[\left\Vert \mathbf{X}^{*}-\mathbf{X}\right\Vert ^{2}\right]\rightarrow0$
as $n\rightarrow\infty$.\end{mydef}

Since $h\left(\mathbf{X}\right)$ is 1/2-strongly convex, we have
$\mathsf{d}_{\textrm{KL}}\left(\mathbf{X}^{*},\mathbf{X}\right)\geq\frac{1}{4}\left\Vert \mathbf{X}^{*}-\mathbf{X}\right\Vert ^{2}$
by the property of Bregman divergence \cite{shalev2012online}, which
means that $D_{n}^{\left(\mathrm{I}\right)}\rightarrow0$ implies
$\mathbb{E}\left[\left\Vert \mathbf{X}^{*}-\mathbf{X}\right\Vert ^{2}\right]\rightarrow0$.
Then we can easily conclude the following result.

\begin{mycor}As long as the condition \eqref{eq:condi_i} holds,
we have $D_{n}^{\left(\mathrm{I}\right)}\rightarrow0$ as $n\rightarrow\infty$,
the MXL algorithm converges in mean square to NE.\end{mycor}

The decreasing order of the average divergence is the same as that
of the step-size $\gamma_{n}$, which does not depend on $p_{\textrm{I}}$.
In fact, the \emph{original} MXL with full gradient knowledge is a
special case with $p_{\textrm{I}}=1$ and $D_{n}^{\left(\textrm{O}\right)}$
can also be bounded by \eqref{eq:dni}.\footnote{Note that the upper bound of $D_{n}^{\left(\textrm{O}\right)}$ presented
in \cite{1} is $D_{n}\leq\lambda'/n$ with $\lambda'$ some constant,
which is derived by considering a special example $\gamma_{n}=\alpha/n$.
Whereas, we consider general setting of the step size $\gamma_{n}$.
In fact, \eqref{eq:dni} turns to be $D_{n}\leq\lambda\alpha/n$ as
$\gamma_{n}=\alpha/n$ and.$p_{\textrm{I}}=1$} It is obvious that the incompleteness of the feedback only affects
the constant term $\lambda$ of the upper bound of $D_{n}^{\left(\mathrm{I}\right)}$,
while the decreasing order only depends on the step-size $\gamma_{n}$. 

Now we consider an example of $\gamma_{n}$ and investigate $\varepsilon$
to show that the condition \eqref{eq:condi_i} can be easily satisfied.

\begin{eg}\label{exa:agmma}In most work related to stochastic approximation,
a common setting of $\gamma_{n}$ is
\begin{equation}
\gamma_{n}=\alpha n^{-\nu},\label{eq:gamma_example}
\end{equation}
with $\nu\in(0.5,1]$ and $\alpha\in\mathbb{R}^{+}$. Here the constraint
on $\nu$ is to make $\gamma_{n}$ satisfy the assumption \emph{A2}.\end{eg}

We check first whether $\varepsilon$ is bounded when $\gamma_{n}$
follows \eqref{eq:gamma_example}. Due to the fact that the denominator
$\gamma_{n}^{2}\rightarrow0$ as $n\rightarrow\infty$, we mainly
need to check 
\begin{align}
\lim_{n\rightarrow\infty}\frac{\gamma_{n}-\gamma_{n+1}}{\gamma_{n}^{2}} & =\lim_{n\rightarrow\infty}\frac{1-\left(1+\frac{1}{n}\right)^{-\nu}}{\alpha n^{-\nu}}=\lim_{x\rightarrow0}\frac{1-\left(1+x\right)^{-\nu}}{\alpha x^{\nu}}\nonumber \\
 & =\lim_{x\rightarrow0}\frac{\left(1+x\right)^{-\nu-1}}{\alpha x^{\nu-1}}=\lim_{x\rightarrow0}\frac{1}{\alpha}x^{1-\nu}\nonumber \\
 & =\begin{cases}
\infty, & \textrm{if }\nu>1,\\
1/\alpha, & \textrm{if }\nu=1,\\
0, & \textrm{else}.
\end{cases}\label{eq:infinit}
\end{align}
Hence $\frac{\gamma_{n}-\gamma_{n+1}}{\gamma_{n}^{2}}$ is bounded
in our case as $\nu\leq1$, we can see clearly the importance of the
assumption \emph{A2}. 

Then we aim to find an upper bound of $\varepsilon$. In fact 
\[
\varepsilon=\max_{n\geq1}\frac{1-\left(1+\frac{1}{n}\right)^{-\nu}}{\alpha n}\leq\max_{0<x\leq1}\frac{g\left(x\right)}{\alpha},
\]
with 
\[
g\left(x\right)=x^{-\nu}\left(1-\left(1+x\right)^{-\nu}\right),
\]
the inequality mainly comes from the fact that $\frac{1}{n}$ takes
discrete values from the set $\left\{ 1,\frac{1}{2},\frac{1}{3},\ldots\right\} $,
while $x$ takes any real values from the interval $(0,1]$. The lemma
in what follows describes an upper bound of $\varepsilon$. 
\begin{lem}
\label{lem:bound_i}For any $x\in\left(0,1\right]$ and $\nu\in\left(0.5,1\right]$,
we have
\begin{equation}
g\left(x\right)\leq\overline{\varepsilon}\left(\nu\right)=\begin{cases}
\nu\left(\frac{1-\nu}{2\nu}\right)^{1-\nu}, & \textrm{if }\nu\in(\log_{2}1.5,1],\\
1-2^{-\nu}, & \textrm{if }\nu\in(0.5,\log_{2}1.5].
\end{cases}\label{eq:gx_bound}
\end{equation}
\end{lem}
\begin{IEEEproof}
See Appendix \ref{subsec:Proof-of-Lemma_boundi}.
\end{IEEEproof}
By applying Lemma \ref{lem:bound_i}, the condition \eqref{eq:condi_i}
holds if $\frac{1}{\alpha}\overline{\varepsilon}\left(\nu\right)<p_{\textrm{I}}B<\frac{1}{\alpha}$.
As $\nu\in\left(0.5,1\right]$, we can verify that $\overline{\varepsilon}\left(\nu\right)\leq1$
with the equality if and only if $\nu=1$. We can conclude that for
any fixed $\nu$, $\alpha$ should be chosen from the interval $\left(\frac{\overline{\varepsilon}\left(\nu\right)}{p_{\textrm{I}}B},\frac{1}{p_{\textrm{I}}B}\right)$. 

For other more complicated forms of $\gamma_{n}$, the conditions
\eqref{eq:condi_i} can also be verified numerically.

\section{Sporadic MXL}

\label{sec:Asynchronous-MXL}

In this section, we introduce and analyze the sporadic version of
the MXL algorithm. 

Different from the situation that all the receivers send incomplete
feedback information, now we consider the scenario that a part of
transmitters send complete feedback information at each time instant.
Particularly, we consider the following setting:

\emph{S2}. At each iteration $n$ of the algorithm, each receiver
$k$ has a probability $p_{\textrm{S}}$ to send the feedback $\widehat{\mathbf{V}}_{k}\left(n\right)$. 

The transmitters update their action only when they have received
the feedback. Let a Bernoulli random variable $\eta_{k}\left(n\right)\in\left\{ 0,1\right\} $
indicate whether receiver~$k$ feeds back the gradient matrix. We
have $\eta_{k}\left(n\right)\sim\mathcal{B}\left(1,p_{\textrm{S}}\right)$
according to the setting \emph{S2}. 

In the sporadic MXL (MXL-S) algorithm, at each iteration~$n\geq1$,
each transmitter~$k$ update their action using
\begin{align}
\mathbf{X}_{k}^{\left(\textrm{S}\right)}\left(n\right) & =A_{k}\frac{\exp\left(\mathbf{Y}_{k}^{\left(\textrm{S}\right)}\left(n-1\right)\right)}{1+\Vert\exp\left(\mathbf{Y}_{k}^{\left(\textrm{S}\right)}\left(n-1\right)\right)\Vert},\label{eq:mxl_s}\\
\mathbf{Y}_{k}^{\left(\textrm{S}\right)}\left(n\right) & =\mathbf{Y}_{k}^{\left(\textrm{S}\right)}\left(n-1\right)+\gamma_{n_{k}}\eta_{k}\left(n\right)\widehat{\mathbf{V}}_{k}\left(n\right),\label{eq:gradient_s}
\end{align}
where $n_{k}$ is maintained by each transmitter $k$ to indicate
the index of the step-size $\gamma_{\cdot}$ to be applied during
algorithm. In this paper, we consider $n_{k}=\widetilde{n}_{k}+\eta_{k}\left(t\right)$,
with $\widetilde{n}_{k}$ a purely Binomial random variable, $\widetilde{n}_{k}\sim\mathcal{B}\left(n-1,p_{\textrm{S}}\right)$. 

The analysis of MXL-S is more complex than that of MXL-I. The main
difference is that: in MXL-S, transmitters update their action at
random time slot and the step-sizes used by different transmitters
are completely independent; while in MXL-I, transmitters update their
action at each time slot and their step-size is always identical. 

The desirable property of $\gamma_{n_{k}}\eta_{k}\left(n\right)$
is stated in the following lemma, which is essential to our main result. 
\begin{lem}
\label{lem:gamma_sum}Denote $\overline{\gamma}_{n}=\mathbb{E}\left[\gamma_{n_{k}}\eta_{k}\left(n\right)\right]$
and $\mathring{\gamma}_{n}=\sqrt{\mathbb{E}\left[\left(\gamma_{n_{k}}\eta_{k}\left(n\right)\right)^{2}\right]}$,
then we have 
\begin{align}
\overline{\gamma}_{n} & =\sum_{\ell=1}^{n}\gamma_{\ell}p_{\mathrm{S}}^{\ell}\left(1-p_{\mathrm{S}}\right)^{n-\ell}\binom{n-1}{\ell-1},\label{eq:ga_1}\\
\mathring{\gamma}_{n}^{2} & =\sum_{\ell=1}^{n}\gamma_{\ell}^{2}p_{\mathrm{S}}^{\ell}\left(1-p_{\mathrm{S}}\right)^{n-\ell}\binom{n-1}{\ell-1}.\label{eq:ga_2}
\end{align}
More importantly, 
\begin{equation}
\sum_{n=1}^{\infty}\overline{\gamma}_{n}=\sum_{n=1}^{\infty}\gamma_{n}\rightarrow\infty\label{eq:gamma_aver_sum}
\end{equation}
\begin{equation}
\sum_{n=1}^{\infty}\overline{\gamma}_{n}^{2}\leq\sum_{n=1}^{\infty}\mathring{\gamma}_{n}^{2}=\sum_{n=1}^{\infty}\gamma_{n}^{2}<\infty\label{eq:gamma_aversquar}
\end{equation}
\end{lem}
\begin{IEEEproof}
See Appendix~\ref{subsec:Proof-of-sum}.
\end{IEEEproof}
In the rest of this section, the convergence of MXL-S is discussed. 

\subsection{\label{subsec:Convergence-of-MXL-S}Convergence of MXL-S}

As in Section~\ref{subsec:Convergence-of-MXL-I}, we assume that
the NE $\mathbf{X}^{*}\in\mathcal{X}$ is globally stable. We investigate
the a.s. convergence of MXL-S, \emph{i.e.}, to check whether $\mathit{d_{n}^{\left(\textrm{S}\right)}}=\mathsf{d}_{\textrm{KL}}\left(\mathbf{X}^{*},\mathbf{X}^{\left(\textrm{S}\right)}\left(n\right)\right)\rightarrow0$
a.s.. 

The main challenge is that the update performed by each transmitter
is at random time slot and the step-size is also generated in a random
manner. Our result is stated in what follows.

\begin{mythm}\label{thm:mxls_conv}Suppose that Assumptions A1-A3
are satisfied, then MXL-S converges to NE a.s., i.e., $\mathit{d_{n}^{\left(\mathrm{S}\right)}}\rightarrow0$
a.s..\end{mythm}
\begin{IEEEproof}
See Appendix~\ref{subsec:Proof-of-conv_mxl_s}.
\end{IEEEproof}
The main idea of the proof is to compare the MXL-S with an original
MXL using the average step-size $\overline{\gamma}_{n}$ defined in
\eqref{eq:ga_1}, which converges to NE a.s. according to the property
of $\overline{\gamma}_{n}$ as stated in Lemma~\ref{lem:gamma_sum}.
Their difference can be seen as another type of stochastic noise to
be learned with the help of Lemma~\ref{lem:gamma_sum}, \emph{i.e.},
the property of $\mathring{\gamma}_{n}$. Note that the tool used
in the proof can be applied to analyze the MXL with asynchronous update
of the transmitter, even if the purpose here is beyond this issue. 

\subsection{Convergence rate of MXL-S}

For the MXL-S algorithm, we use $D_{n}^{\left(\textrm{S}\right)}$
to denote the average quantum KL divergence. We re-consider Assumption~A4
to get the result as presented in Theorem~\ref{prop:rate_asy}.

\begin{mythm}\label{prop:rate_asy}Assume that the assumptions A1,
A2, and A4 hold, the MXL-S algorithm is performed such that each receiver
feeds back with probability $p_{\mathrm{S}}$. Let 
\begin{equation}
\epsilon=\max\frac{1}{\mathring{\gamma}_{n}^{2}}\left(\frac{\mathring{\gamma}_{n}^{2}}{\overline{\gamma}_{n}}-\frac{\mathring{\gamma}_{n+1}^{2}}{\overline{\gamma}_{n+1}}\right),
\end{equation}
then if $\epsilon<B<\frac{1}{\gamma_{1}}$, we have 
\begin{equation}
D_{n}^{\left(\mathrm{S}\right)}\leq\mu\frac{\mathring{\gamma}_{n}^{2}}{\overline{\gamma}_{n}}\textrm{ }with\textrm{ }\mu=\max\left\{ \frac{D_{1}^{\left(\mathrm{S}\right)}}{\gamma_{1}},\frac{C}{B-\epsilon}\right\} .
\end{equation}
\end{mythm}
\begin{IEEEproof}
See Appendix~\ref{subsec:Proof-2}. 
\end{IEEEproof}
Compared with $D_{n}^{\left(\mathrm{I}\right)}$, the decreasing order
of $D_{n}^{\left(\mathrm{S}\right)}$ is more complicated , as it
depends on the equivalent step-size $\mathring{\gamma}_{n}^{2}/\overline{\gamma}_{n}$
which is affected by $p_{\textrm{S}}$. We have obtained an upper
bound of $\mathring{\gamma}_{n}^{2}/\overline{\gamma}_{n}$ described
in the following lemma, which can be used to prove the convergence
of the sporadic MXL to NE from a theoretical point of view. 
\begin{lem}
\label{lem:rate_bound}If $\gamma_{n}$ is convex over $n$, then
for any constant $\xi\in\left(0,1\right)$ we have\textup{
\begin{equation}
\frac{\mathring{\gamma}_{n+1}^{2}}{\overline{\gamma}_{n+1}}\leq\frac{\exp\left(-\frac{1}{2}\xi^{2}p_{\mathrm{S}}n\right)\gamma_{1}^{2}+\gamma_{\left\lfloor \left(1-\xi\right)p_{\mathrm{S}}n\right\rfloor +2}^{2}}{\gamma_{\left\lfloor p_{\mathrm{S}}n\right\rfloor +2}}.\label{eq:rate_a}
\end{equation}
}
\end{lem}
\begin{IEEEproof}
See Appendix \ref{subsec:Proof-of-Lemma_bound}.
\end{IEEEproof}
Note that the convexity of $\gamma_{n}$ is not a big assumption as
$\gamma_{n}$ satisfies \emph{A2}. In fact, $\gamma_{n}=\alpha n^{-\nu}$
as introduced in Example \ref{exa:agmma} is convex. 

\begin{mycor}If $\gamma_{n}=\alpha n^{-\nu}$ with $\alpha\in\mathbb{R}^{+}$
and $\nu\in(0.5,1]$, then $D_{n}^{\left(\mathrm{S}\right)}\rightarrow0$
as $n\rightarrow\infty$, the MXL algorithm converges in mean square
to NE.\end{mycor}
\begin{IEEEproof}
From Lemma \ref{lem:rate_bound}, we can evaluate 
\begin{align}
 & \lim_{n\rightarrow\infty}\frac{\mathring{\gamma}_{n+1}^{2}}{\overline{\gamma}_{n+1}}\nonumber \\
 & \leq\lim_{n\rightarrow\infty}\alpha\frac{\exp\left(-\frac{1}{2}\xi^{2}p_{\textrm{S}}n\right)+\left(\left\lfloor \left(1-\xi\right)p_{\textrm{S}}n\right\rfloor +2\right)^{-2\nu}}{\left(\left\lfloor p_{\textrm{S}}n\right\rfloor +2\right)^{-\nu}}\nonumber \\
 & =\lim_{n\rightarrow\infty}\alpha\left(\frac{\left\lfloor p_{\textrm{S}}n\right\rfloor +2}{\left(\left\lfloor \left(1-\xi\right)p_{\textrm{S}}n\right\rfloor +2\right)^{2}}\right)^{\nu}\nonumber \\
 & =\lim_{n\rightarrow\infty}\alpha\left(\left(1-\xi\right)^{2}p_{\textrm{S}}\right)^{-\nu}n^{-\nu}=0,\label{eq:gamma_equiv}
\end{align}
therefore $\mathring{\gamma}_{n}^{2}/\overline{\gamma}_{n}$ is vanishing
when $\gamma_{n}=\alpha n^{-\nu}$. In fact, we can see that $\mathring{\gamma}_{n}^{2}/\overline{\gamma}_{n}\propto n^{-\nu}$
as $n$ is large enough from \eqref{eq:gamma_equiv}. Besides, the
constant term $\mu$ is bounded, we can then conclude that $D_{n}^{\left(\mathrm{S}\right)}\rightarrow0$
as $n\rightarrow\infty$, which implies the convergence of the MXL
algorithm to NE as $\gamma_{n}=\alpha n^{-\nu}$.
\end{IEEEproof}

\section{Numerical Example}

\label{sec:Numerical-Example}

In our simulation, we consider $K=9$ pairs of transmitter-receiver
links. For each link, we set $P_{\textrm{c}}=20\textrm{dBm}$, $P_{\textrm{max}}=30\textrm{dBm}$,
$N_{t}=4$, $N_{r}=8$ and $S=3$. The additive noise $\mathbf{Z}$
is generated as Gaussian random variable with zero-mean and variance
$1$. For each different setting, 100 independent simulations are
preformed to obtain the average results. 

In the MXL algorithm, we set the initial transmit power as $P_{\textrm{max}}/2$
and the step-size is $\gamma_{n}=0.2n^{-0.7}$. For short, we use
MXL-I and MXL-S to name the MXL algorithm with incomplete feedback
and the sporadic MXL, respectively. 

\begin{figure}
\begin{centering}
\includegraphics[width=0.97\columnwidth]{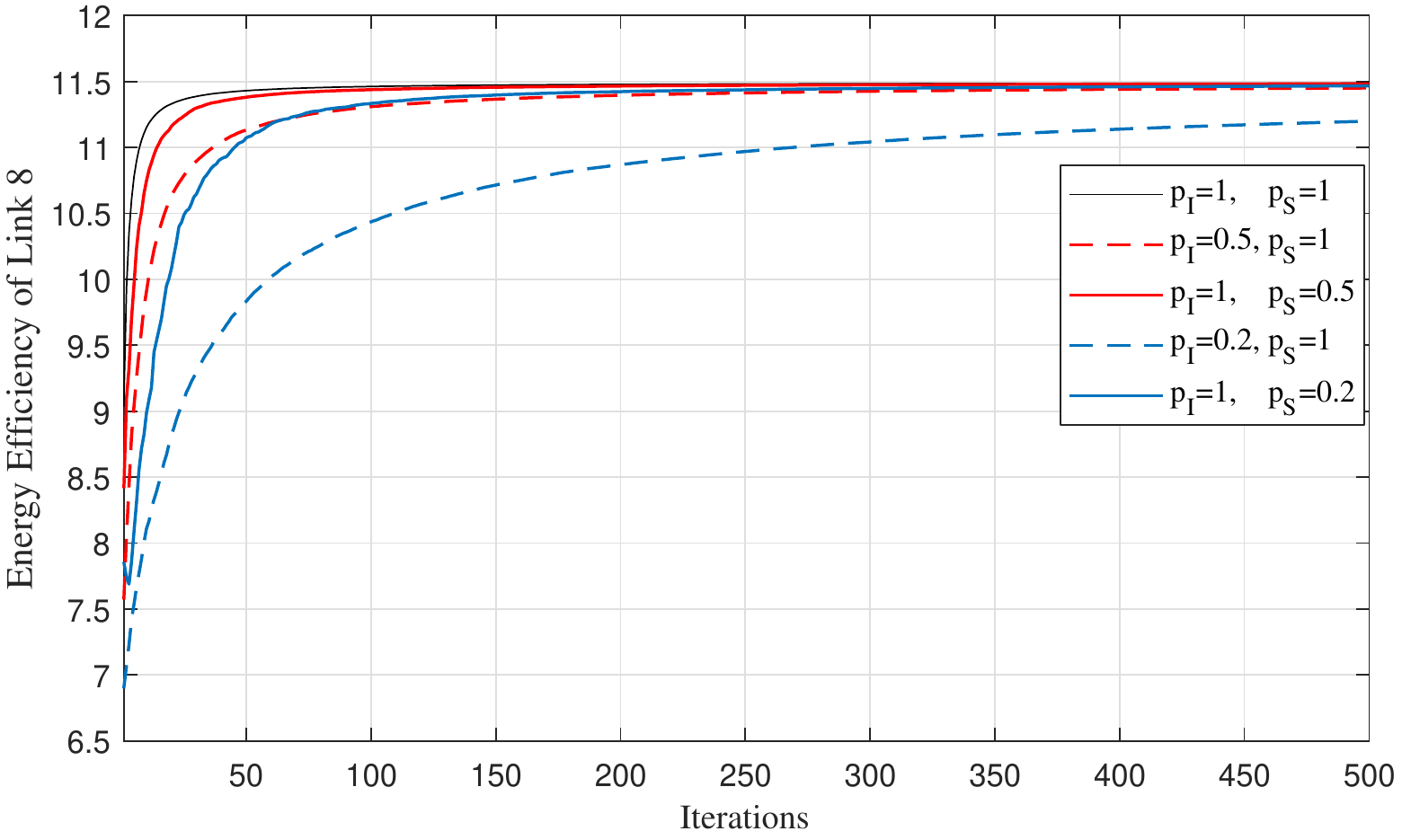}
\par\end{centering}
\caption{\label{fig:EE1}For a static channel, evolution of energy efficiency
of link 8, average results from 100 simulations by performing: (i)
MXL-I with $p_{\textrm{I}}\in\left\{ 0.2,0.5\right\} $ and $p_{\textrm{S}}=1$;
(ii) MXL-S with $p_{\textrm{S}}\in\left\{ 0.2,0.5\right\} $ and $p_{\textrm{I}}=1$;
(iii) original MXL with $p_{\textrm{I}}=p_{\textrm{S}}=1$. }
\end{figure}

\begin{figure}
\begin{centering}
\includegraphics[width=0.97\columnwidth]{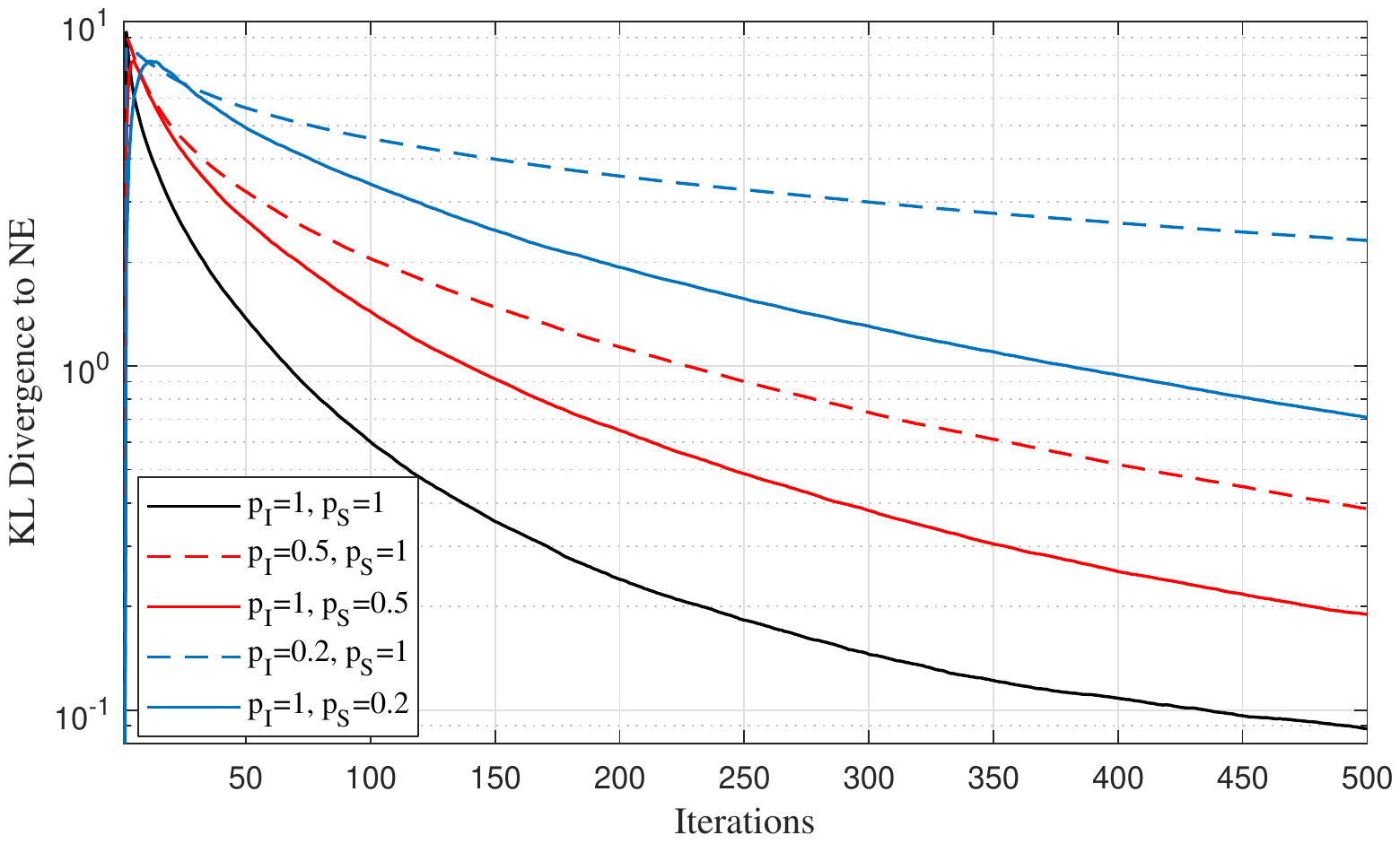}
\par\end{centering}
\caption{\label{fig:diver1}For a static channel, evolution of the average
quantum Kullback-Leibler divergence to NE, average results from 100
simulations by performing: (i) MXL-I with $p_{\textrm{I}}\in\left\{ 0.2,0.5\right\} $
and $p_{\textrm{S}}=1$; (ii) MXL-S with $p_{\textrm{S}}\in\left\{ 0.2,0.5\right\} $
and $p_{\textrm{I}}=1$; (iii) original MXL with $p_{\textrm{I}}=p_{\textrm{S}}=1$. }
\end{figure}

We start with an easy situation: the channel matrix keeps \emph{static}
with its initial value randomly generated. The evolution of the EE
of different links has similar shape. Consider an arbitrary link 8,
Figure~\ref{fig:EE1} compares the average evolution of $u_{8}$
(EE) obtained by performing MXL-I with $p_{\textrm{I}}\in\left\{ 0.2,0.5\right\} $
and $p_{\textrm{S}}=1$, MXL-S with $p_{\textrm{S}}\in\left\{ 0.2,0.5\right\} $
and $p_{\textrm{I}}=1$, as well as the original MXL with $p_{\textrm{I}}=p_{\textrm{S}}=1$.
Furthermore, Figure~\ref{fig:diver1} shows the average divergence
$D_{n}^{\left(\textrm{I}\right)}$ and $D_{n}^{\left(\textrm{S}\right)}$
in all these cases. We can see that the MXL algorithm tends to converge
to NE in all the cases. For the same level of traffic, for example,
MXL-I with $p_{\textrm{I}}=0.5$ and MXL-S with $p_{\textrm{S}}=0.5$,
we find that MXL-S converges faster than MXL-I. In fact, MXL-S converges
slightly slower compared with the original MXL, even if half of the
signaling information is reduced. Another interesting result is that
the performance of MXL-S is less sensitive to $p_{\textrm{S}}$, while
MXL-I is more sensitive to $p_{\textrm{I}}$. As we can see from the
results, the difference between the curves related to MXL-S with $p_{\textrm{S}}=0.5$
and $p_{\textrm{S}}=0.2$ are much smaller than the difference presented
in MXL-I with $p_{\textrm{I}}=0.5$ and $p_{\textrm{I}}=0.2$. 


\begin{figure}
\begin{centering}
\includegraphics[width=0.97\columnwidth]{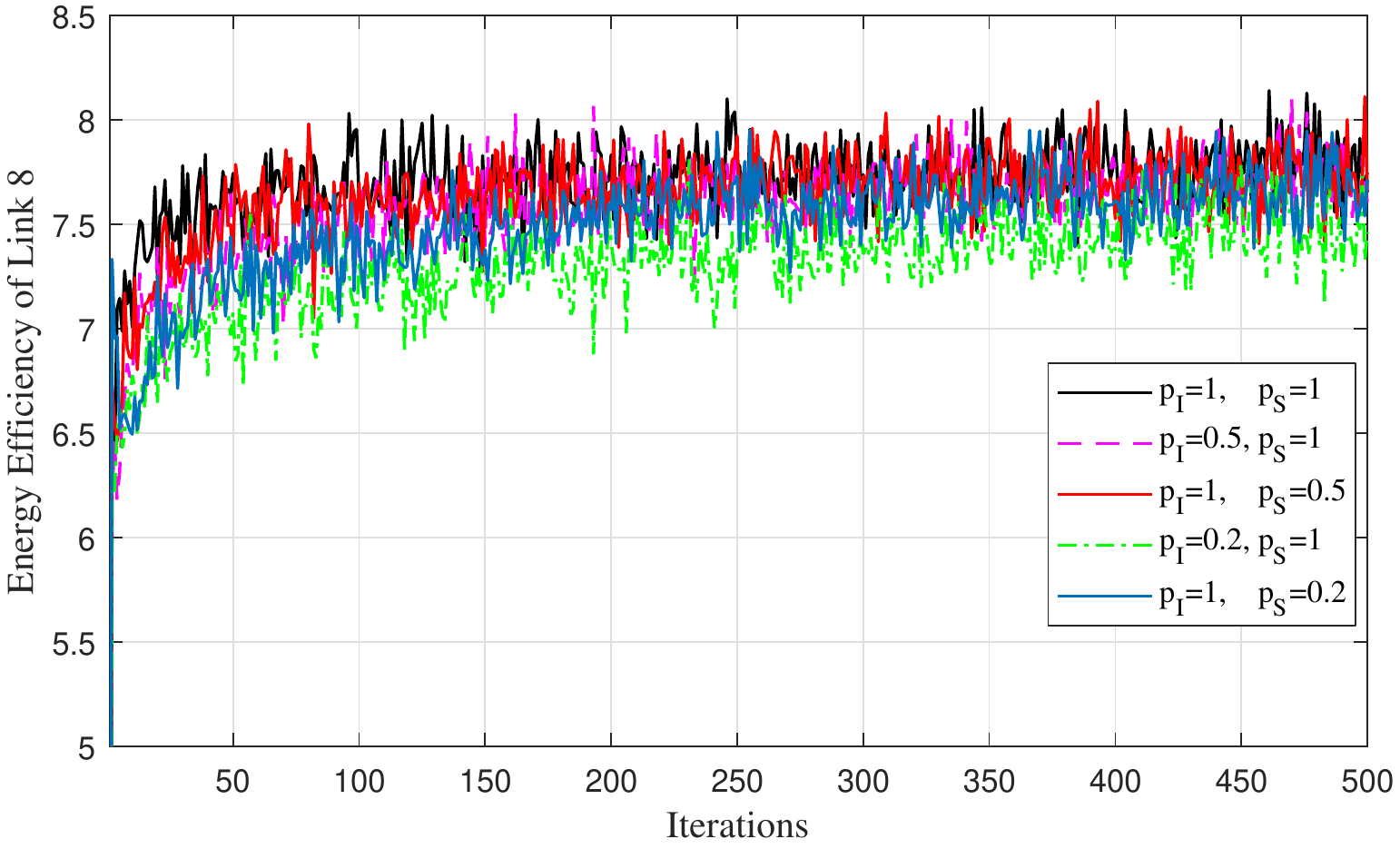}
\par\end{centering}
\caption{\label{fig:EE2}For a stochastic channel, evolution of energy efficiency
of link 8, average results from 100 simulations by performing: (i)
MXL-I with $p_{\textrm{I}}\in\left\{ 0.2,0.5\right\} $ and $p_{\textrm{S}}=1$;
(ii) MXL-S with $p_{\textrm{S}}\in\left\{ 0.2,0.5\right\} $ and $p_{\textrm{I}}=1$;
(iii) original MXL with $p_{\textrm{I}}=p_{\textrm{S}}=1$. }
\end{figure}

\begin{figure}
\begin{centering}
\includegraphics[width=0.97\columnwidth]{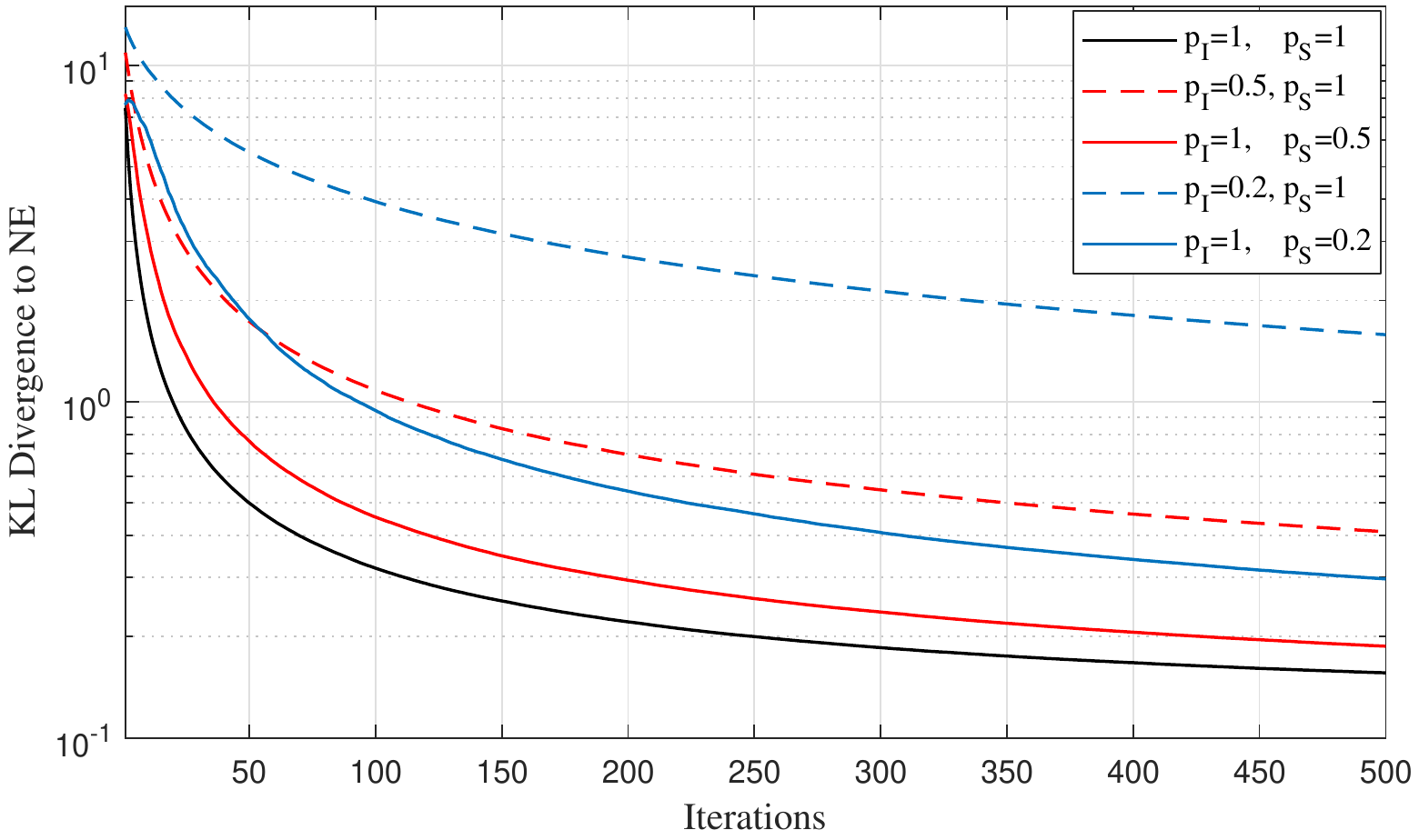}
\par\end{centering}
\caption{\label{fig:diver2}For a stochastic channel, evolution of the average
quantum Kullback-Leibler divergence to NE, average results from 100
simulations by performing: (i) MXL-I with $p_{\textrm{I}}\in\left\{ 0.2,0.5\right\} $
and $p_{\textrm{S}}=1$; (ii) MXL-S with $p_{\textrm{S}}\in\left\{ 0.2,0.5\right\} $
and $p_{\textrm{I}}=1$; (iii) original MXL with $p_{\textrm{I}}=p_{\textrm{S}}=1$. }
\end{figure}

Then we consider a much more challenging situation where channel matrix
is stochastic and its elements are randomly and independently generated
at each iteration. Results are presented in Figures~\ref{fig:EE2}
and~\ref{fig:diver2}, which are similar compared with Figures~\ref{fig:EE1}
and~\ref{fig:diver1}. We can see that average EE is quite sensitive
to the stochastic channel, while the evolution $D_{n}$ is smooth
by the average of 100 simulations. Our claim is thus further justified
by the Figures~\ref{fig:EE2} and~\ref{fig:diver2}.


\section{Conclusion}

In this paper, we investigate the performance of the MXL algorithm
under different feedback strategies. We have proposed two variants
of the MXL algorithm in order to reduce the signaling overhead: one
is by making receivers feedback only part the elements of the gradient
matrix per iteration; the other is by making receivers sporadically
feedback the whole gradient matrix. For both strategies, we have proved
the convergence of the MXL algorithm to NE and evaluated the upper
bounds of the average convergence rate as well. From the theoretical
results we can clearly see that the incompleteness of the feedback
information does not seriously affect the convergence rate of the
MXL algorithm. In the simulations, we consider a distributed energy
efficiency maximization problem in a multi-user, multicarrier MIMO
network. The results are provided to justify our claim. In some scenario,
such as the simulation that we considered, the second proposed strategy
performs better in terms of the convergence rate.

\label{sec:Conclusion}

\appendix

\subsection{\label{subsec:Proof-pre}Proof sketch of Lemma~\ref{lem:MXL_pre}}

We present a brief proof that has been presented in \cite{1}. Consider
\begin{align}
h^{*}\left(\mathbf{Y}\right) & =\max_{\mathbf{X}\in\mathbb{H}_{+}^{M}:\textrm{tr}\left(\mathbf{X}\right)\leq1}\left[\textrm{tr}\left(\mathbf{Y}\mathbf{X}\right)-h\left(\mathbf{X}\right)\right],\label{eq:hstar}
\end{align}
where $h^{*}$ denotes the convex conjugate of $h$ over a spectrahedron,
with $\mathbf{Y}\in\mathbb{H}^{M}$. As stated in Proposition A.1
of \cite{1}, the closed expression of $h^{*}\left(\mathbf{Y}\right)$
can be derived, \emph{i.e.}, $h^{*}\left(\mathbf{Y}\right)=\log\left(1+\textrm{tr}\left(\exp\left(\mathbf{Y}\right)\right)\right)$.
Introduce 
\begin{equation}
G\left(\mathbf{Y}\right)=\frac{\exp\left(\mathbf{Y}\right)}{1+\textrm{tr}\left(\exp\left(\mathbf{Y}\right)\right)},\label{eq:gy}
\end{equation}
then it is straightforward to deduce that $G\left(\mathbf{Y}\right)$
is the gradient matrix of $h^{*}\left(\mathbf{Y}\right)$, \emph{i.e.},
$G\left(\mathbf{Y}\right)=\nabla h^{*}\left(\mathbf{Y}\right)$. 

With the above definition, for the NE $\mathbf{X}^{*}$ and any $\mathbf{Y}\in\mathbb{H}^{M}$,
we consider the Fenchel coupling
\begin{equation}
F\left(\mathbf{X}^{*},\mathbf{Y}\right)=h\left(\mathbf{X}^{*}\right)+h^{*}\left(\mathbf{Y}\right)-\textrm{tr}\left(\mathbf{Y}\mathbf{X}^{*}\right).\label{eq:fenchel}
\end{equation}
We have $F\left(\mathbf{X}^{*},\mathbf{Y}\right)\geq0$ by Fenchel\textendash Young
inequality, with equality iif $\mathbf{X}^{*}=\nabla h^{*}\left(\mathbf{Y}\right)=G\left(\mathbf{Y}\right)$.
According to the equivalence between Fenchel coupling and KL divergence
presented in \cite{mertikopoulos2016learning}, we have $\mathsf{d}_{\mathrm{KL}}\left(\mathbf{X}^{*},G\left(\mathbf{Y}\right)\right)=F\left(\mathbf{X}^{*},\mathbf{Y}\right)$.

In the MXL algorithm, we find that \eqref{eq:mxl}can be written as
$\mathbf{X}\left(n\right)=G\left(\mathbf{Y}\left(n\right)\right)$.
Hence, 
\begin{align}
 & \mathsf{d}_{\mathrm{KL}}\left(\mathbf{X}^{*},\mathbf{X}\left(n+1\right)\right)=\mathsf{d}_{\mathrm{KL}}\left(\mathbf{X}^{*},G\left(\mathbf{Y}\left(n+1\right)\right)\right)\nonumber \\
 & =F\left(\mathbf{X}^{*},\mathbf{Y}\left(n+1\right)\right)\overset{\left(a\right)}{=}F\left(\mathbf{X}^{*},\mathbf{Y}\left(n\right)+\mathbf{U}\left(n\right)\right)\nonumber \\
 & \overset{\left(b\right)}{\leq}F\left(\mathbf{X}^{*},\mathbf{Y}\left(n\right)\right)+\textrm{tr}\left(\left(G\left(\mathbf{Y}\left(n\right)\right)-\mathbf{X}^{*}\right)\mathbf{U}\left(n\right)\right)+\left\Vert \mathbf{U}\left(n\right)\right\Vert _{\infty}^{2}\nonumber \\
 & =\mathsf{d}_{\mathrm{KL}}\left(\mathbf{X}^{*},\mathbf{X}\left(n\right)\right)+\textrm{tr}\left(\left(\mathbf{X}\left(n\right)-\mathbf{X}^{*}\right)\mathbf{U}\left(n\right)\right)+\left\Vert \mathbf{U}\left(n\right)\right\Vert _{\infty}^{2},\label{eq:d_basic}
\end{align}
where we introduce $\mathbf{U}\left(n\right)=\mathbf{Y}\left(n+1\right)-\mathbf{Y}\left(n\right)$
in $\left(a\right)$, recall that $\mathbf{U}\left(n\right)=\gamma_{n}\widehat{\mathbf{V}}\left(n\right)$
in the MXL algorithm, according to \eqref{eq:gradient_descent}; $\left(b\right)$
is by Proposition A.2 in \cite{1}, which can be proved by the strong
smoothness of $h^{*}$.

\subsection{\label{subsec:Proof-convmxli}Proof of Theorem~\ref{thm:mxli_conv}}

Define
\[
d_{n}^{\left(\textrm{I}\right)}=\mathsf{d}_{\mathrm{KL}}\left(\mathbf{X}^{*},\mathbf{X}^{\left(\textrm{I}\right)}\left(n\right)\right).
\]
According to Lemma~\ref{lem:MXL_pre}, we have
\begin{align}
d_{n+1}^{\left(\textrm{I}\right)} & \leq d_{n}^{\left(\textrm{I}\right)}+\gamma_{n}\textrm{tr}\left(\left(\mathbf{X}^{\left(\textrm{I}\right)}\left(n\right)-\mathbf{X}^{*}\right)\boldsymbol{\Delta}\left(n\right)\circ\widehat{\mathbf{V}}\left(n\right)\right)\nonumber \\
 & \qquad\qquad+\gamma_{n}^{2}\left\Vert \boldsymbol{\Delta}\left(n\right)\circ\widehat{\mathbf{V}}\left(n\right)\right\Vert _{\infty}^{2},\label{eq:dn+1_n_mxli}
\end{align}
by the fact that $\mathbf{U}\left(n\right)=\gamma_{n}\boldsymbol{\Delta}\left(n\right)\circ\widehat{\mathbf{V}}\left(n\right)$
in MXL-I according to \eqref{eq:gradient_i}. Note that the presence
of $\boldsymbol{\Delta}\left(n\right)$ makes the proof more complicated,
compared with that in \cite{1}. Introduce
\begin{align}
 & e_{\boldsymbol{\Delta}}\left(n\right)=\textrm{tr}\left(\left(\mathbf{X}^{\left(\textrm{I}\right)}\left(n\right)-\mathbf{X}^{*}\right)\boldsymbol{\Delta}\left(n\right)\circ\widehat{\mathbf{V}}\left(n\right)\right)\nonumber \\
 & \qquad-\mathbb{E}_{\boldsymbol{\Delta}}\left[\textrm{tr}\left(\left(\mathbf{X}^{\left(\textrm{I}\right)}\left(n\right)-\mathbf{X}^{*}\right)\boldsymbol{\Delta}\left(n\right)\circ\widehat{\mathbf{V}}\left(n\right)\right)\right].\label{eq:e_delta}
\end{align}
Since the elements of $\boldsymbol{\Delta}\left(n\right)$ follow
i.i.d. Bernoulli distribution, it is easy to obtain $\mathbb{E}_{\boldsymbol{\Delta}}\left(\boldsymbol{\Delta}\left(n\right)\circ\widehat{\mathbf{V}}\left(n\right)\right)=p_{\mathrm{I}}\widehat{\mathbf{V}}\left(n\right)$,
hence 
\begin{align}
 & \mathbb{E}_{\boldsymbol{\Delta}}\left[\textrm{tr}\left(\left(\mathbf{X}^{\left(\textrm{I}\right)}\left(n\right)-\mathbf{X}^{*}\right)\boldsymbol{\Delta}\left(n\right)\circ\widehat{\mathbf{V}}\left(n\right)\right)\right]\nonumber \\
 & =p_{\mathrm{I}}\textrm{tr}\left(\left(\mathbf{X}^{\left(\textrm{I}\right)}\left(n\right)-\mathbf{X}^{*}\right)\widehat{\mathbf{V}}\left(n\right)\right).\label{eq:delta_aver}
\end{align}
Similar to \eqref{eq:e_delta}, we define another stochastic noise
arisen by the the approximation of the gradient matrix, 
\begin{align}
e_{\mathbf{Z}}^{\left(\mathrm{I}\right)}\left(n\right) & =p_{\mathrm{I}}\textrm{tr}\left(\left(\mathbf{X}^{\left(\textrm{I}\right)}\left(n\right)-\mathbf{X}^{*}\right)\widehat{\mathbf{V}}\left(n\right)\right)\nonumber \\
 & -p_{\mathrm{I}}\mathbb{E}_{\mathbf{Z}}\left[\textrm{tr}\left(\left(\mathbf{X}^{\left(\textrm{I}\right)}\left(n\right)-\mathbf{X}^{*}\right)\left(\mathbf{V}\left(n\right)+\mathbf{Z}\left(n\right)\right)\right)\right]\nonumber \\
 & =p_{\mathrm{I}}\textrm{tr}\left(\left(\mathbf{X}^{\left(\textrm{I}\right)}\left(n\right)-\mathbf{X}^{*}\right)\left(\widehat{\mathbf{V}}\left(n\right)-\mathbf{V}\left(n\right)\right)\right)\label{eq:e_z_i}\\
 & =p_{\mathrm{I}}\textrm{tr}\left(\left(\mathbf{X}^{\left(\textrm{I}\right)}\left(n\right)-\mathbf{X}^{*}\right)\mathbf{Z}\left(n\right)\right),\label{eq:e_z_i_1}
\end{align}
which comes from the assumption that the additive noise $\mathbf{Z}$
has zero mean.

Using equations \eqref{eq:e_delta}-\eqref{eq:e_z_i_1}, we evaluate
\begin{align}
 & \textrm{tr}\left(\left(\mathbf{X}^{\left(\textrm{I}\right)}\left(n\right)-\mathbf{X}^{*}\right)\boldsymbol{\Delta}\left(n\right)\circ\widehat{\mathbf{V}}\left(n\right)\right)\nonumber \\
 & =\mathbb{E}_{\boldsymbol{\Delta}}\left[\textrm{tr}\left(\left(\mathbf{X}^{\left(\textrm{I}\right)}\left(n\right)-\mathbf{X}^{*}\right)\boldsymbol{\Delta}\left(n\right)\circ\widehat{\mathbf{V}}\left(n\right)\right)\right]+e_{\boldsymbol{\Delta}}\left(n\right)\nonumber \\
 & =p_{\mathrm{I}}\textrm{tr}\left(\left(\mathbf{X}^{\left(\textrm{I}\right)}\left(n\right)-\mathbf{X}^{*}\right)\widehat{\mathbf{V}}\left(n\right)\right)+e_{\boldsymbol{\Delta}}\left(n\right)\nonumber \\
 & =p_{\mathrm{I}}\textrm{tr}\left(\left(\mathbf{X}^{\left(\textrm{I}\right)}\left(n\right)-\mathbf{X}^{*}\right)\mathbf{V}\left(n\right)\right)+e_{\boldsymbol{\Delta}}\left(n\right)+e_{\mathbf{Z}}^{\left(\mathrm{I}\right)}\left(n\right).\label{eq:step_i1}
\end{align}
Combine \eqref{eq:dn+1_n_mxli} and \eqref{eq:step_i1}, we have 

\begin{align}
 & d_{n+1}^{\left(\textrm{I}\right)}\leq d_{n}^{\left(\textrm{I}\right)}+p_{\mathrm{I}}\gamma_{n}\textrm{tr}\left(\left(\mathbf{X}^{\left(\textrm{I}\right)}\left(n\right)-\mathbf{X}^{*}\right)\mathbf{V}\left(n\right)\right)\nonumber \\
 & \quad+\gamma_{n}e_{\boldsymbol{\Delta}}\left(n\right)+\gamma_{n}e_{\mathbf{Z}}^{\left(\mathrm{I}\right)}\left(n\right)+\gamma_{n}^{2}\left\Vert \boldsymbol{\Delta}\left(n\right)\circ\widehat{\mathbf{V}}\left(n\right)\right\Vert _{\infty}^{2}.\label{eq:dn+1_n_mxli-1}
\end{align}
Perform the sum of \eqref{eq:dn+1_n_mxli-1}, we get
\begin{align}
 & d_{N+1}^{\left(\textrm{I}\right)}\leq d_{1}^{\left(\textrm{I}\right)}+p_{\mathrm{I}}\sum_{n=1}^{N}\gamma_{n}\textrm{tr}\left(\left(\mathbf{X}^{\left(\textrm{I}\right)}\left(n\right)-\mathbf{X}^{*}\right)\mathbf{V}\left(n\right)\right)+E_{N}^{\left(\mathrm{I}\right)},\label{eq:dn+1_n_mxli-2}
\end{align}
in which 
\begin{equation}
E_{N}^{\left(\mathrm{I}\right)}=\sum_{n=1}^{N}\gamma_{n}\left(e_{\boldsymbol{\Delta}}\left(n\right)+e_{\mathbf{Z}}^{\left(\mathrm{I}\right)}\left(n\right)+\gamma_{n}\left\Vert \boldsymbol{\Delta}\left(n\right)\circ\widehat{\mathbf{V}}\left(n\right)\right\Vert _{\infty}^{2}\right).\label{eq:esum}
\end{equation}
We introduce an important lemma with the proof presented in Appendix~\ref{subsec:Proof_sto_e1}.
\begin{lem}
\label{lem:bound_E}As long as Assumptions A1-A2 hold, we have $\left|E_{N}^{\left(\mathrm{I}\right)}\right|<\infty$
a.s..
\end{lem}
The rest part of the proof is straightforward and similar to that
in \cite{1}. Recall that $\textrm{tr}\left(\left(\mathbf{X}\left(n\right)-\mathbf{X}^{*}\right)\mathbf{V}\left(n\right)\right)\leq0$
by Assumption \emph{A3}. The basic idea is to suppose that there exists
a small positive constant $c$ and sufficient large $N_{0}$ such
that 
\begin{equation}
\textrm{tr}\left(\left(\mathbf{X}\left(n\right)-\mathbf{X}^{*}\right)\mathbf{V}\left(n\right)\right)\leq-c,\quad\forall n\geq N_{0},\label{eq:hypothsis}
\end{equation}
which leads to
\begin{align}
\lim_{N\rightarrow\infty}\sum_{n=N_{0}}^{N}\gamma_{n}\textrm{tr}\left(\left(\mathbf{X}^{\left(\textrm{I}\right)}\left(n\right)-\mathbf{X}^{*}\right)\mathbf{V}\left(n\right)\right) & \leq-c\sum_{n=N_{0}}^{\infty}\gamma_{n}\nonumber \\
 & <-\infty.
\end{align}
Meanwhile, with $\left|E_{N}^{\left(\mathrm{I}\right)}\right|<\infty$
a.s., we finally get that $\lim_{N\rightarrow\infty}d_{N+1}^{\left(\mathrm{I}\right)}<-\infty$,
which obviously violates the fact that $d_{N+1}^{\left(\mathrm{I}\right)}\geq0$.
Therefore, the hypothesis \eqref{eq:hypothsis} does not hold. Therefore,
we can say that $\mathbf{X}_{k}\left(n\right)$ converges to $\mathbf{X}_{k}^{*}$,
$\forall k$.

\subsection{\label{subsec:Proof_sto_e1}Proof of Lemma~\ref{lem:bound_E}}

In order to prove Lemma~\ref{lem:bound_E}, it is sufficient to show
the following: 
\begin{align}
\lim_{N\rightarrow\infty}\sum_{n=1}^{N}\gamma_{n}^{2}\left\Vert \boldsymbol{\Delta}\left(n\right)\circ\widehat{\mathbf{V}}\left(n\right)\right\Vert _{\infty}^{2} & <\infty,\label{eq:e_1}\\
\lim_{N\rightarrow\infty}\left|\sum_{n=1}^{N}\gamma_{n}e_{\boldsymbol{\Delta}}\left(n\right)\right| & <\infty,\quad\mathrm{a.s.}\label{eq:e_1-1}\\
\lim_{N\rightarrow\infty}\left|\sum_{n=1}^{N}\gamma_{n}e_{\mathbf{Z}}^{\left(\mathrm{I}\right)}\left(n\right)\right| & <\infty,\quad\mathrm{a.s.}\label{eq:e_1-1-1}
\end{align}
We show them separately in this appendix. 

\subsubsection{Proof of \eqref{eq:e_1}}

All the elements of the gradient matrix $\widehat{\mathbf{V}}_{k}$
should have bounded value, since they are approximated by each receiver
and then have to be transmitted within feedback packets. We can say
that there exist a constant $C<\infty$ such that 
\begin{equation}
\sum_{i,j}\left|\widehat{V}_{i,j}\left(n\right)\right|^{2}\leq C.\label{eq:bound_V}
\end{equation}
Hence, we have 

\begin{align}
 & \lim_{N\rightarrow\infty}\sum_{n=1}^{N}\gamma_{n}^{2}\left\Vert \boldsymbol{\Delta}\left(n\right)\circ\widehat{\mathbf{V}}\left(n\right)\right\Vert _{\infty}^{2}\nonumber \\
 & \leq\lim_{N\rightarrow\infty}\sum_{n=1}^{N}\gamma_{n}^{2}\sum_{i,j}\left|\delta_{i,j}\left(n\right)\widehat{V}_{i,j}\left(n\right)\right|^{2}\nonumber \\
 & \leq\lim_{N\rightarrow\infty}\sum_{n=1}^{N}\gamma_{n}^{2}\sum_{i,j}\left|\widehat{V}_{i,j}\left(n\right)\right|^{2}\nonumber \\
 & \leq\lim_{N\rightarrow\infty}C\sum_{n=1}^{N}\gamma_{n}^{2}<\infty,
\end{align}
recall that $\delta_{i,j}\left(n\right)\in\left\{ 0,1\right\} $. 

\subsubsection{Proof of \eqref{eq:e_1-1} and \eqref{eq:e_1-1-1}}

By definition \eqref{eq:e_delta}, it is obvious that: i). $\mathbb{E}_{\boldsymbol{\Delta}}\left[e_{\boldsymbol{\Delta}}\left(n\right)\right]=0$;
ii). $e_{\boldsymbol{\Delta}}\left(n_{1}\right)$ is independent of
$e_{\boldsymbol{\Delta}}\left(n_{2}\right)$ due to the independence
of $\boldsymbol{\Delta}\left(n_{1}\right)$ and $\boldsymbol{\Delta}\left(n_{2}\right)$.
Therefore, $\sum_{n=1}^{N}\gamma_{n}e_{\boldsymbol{\Delta}}\left(n\right)$
is martingale, so that we can use Doob's inequality, to have, for
any $\rho>0$,
\begin{align}
 & \mathbb{P}\left[\sup_{N}\left|\sum_{n=1}^{N}\gamma_{n}e_{\boldsymbol{\Delta}}\left(n\right)\right|\geq\rho\right]\nonumber \\
 & \leq\rho^{-2}\mathbb{E}_{\boldsymbol{\Delta}}\left[\left|\sum_{n=1}^{N}\gamma_{n}e_{\boldsymbol{\Delta}}\left(n\right)\right|^{2}\right]\nonumber \\
 & \overset{\left(a\right)}{=}\rho^{-2}\sum_{n=1}^{N}\gamma_{n}^{2}\mathbb{E}_{\boldsymbol{\Delta}}\left[e_{\boldsymbol{\Delta}}^{2}\left(n\right)\right]\nonumber \\
 & \leq\rho^{-2}\sum_{n=1}^{N}\gamma_{n}^{2}\mathbb{E}_{\boldsymbol{\Delta}}\left[\left(\textrm{tr}\left(\left(\mathbf{X}^{\left(\textrm{I}\right)}\left(n\right)-\mathbf{X}^{*}\right)\boldsymbol{\Delta}\left(n\right)\circ\widehat{\mathbf{V}}\left(n\right)\right)\right)^{2}\right]\nonumber \\
 & \overset{\left(b\right)}{\leq}\rho^{-2}c_{1}\sum_{n=1}^{N}\gamma_{n}^{2},\label{eq:st_f1}
\end{align}
in which: $(a)$ is by the fact that $\mathbb{E}_{\boldsymbol{\Delta}}\left[e_{\boldsymbol{\Delta}}\left(n_{1}\right)e_{\boldsymbol{\Delta}}\left(n_{2}\right)\right]=0$
for any $n_{1}\neq n_{2}$; we have $\left|\textrm{tr}\left(\left(\mathbf{X}^{\left(\textrm{I}\right)}\left(n\right)-\mathbf{X}^{*}\right)\right)\right|<\infty$
and $\left|\textrm{tr}\left(\boldsymbol{\Delta}\left(n\right)\circ\widehat{\mathbf{V}}\left(n\right)\right)\right|<\infty$
as $\widehat{V}_{i,j}\left(n\right)$ takes bounded value, there should
exists $c_{1}<\infty$ such that 
\begin{equation}
\mathbb{E}_{\boldsymbol{\Delta}}\left[\left(\textrm{tr}\left(\left(\mathbf{X}^{\left(\textrm{I}\right)}\left(n\right)-\mathbf{X}^{*}\right)\boldsymbol{\Delta}\left(n\right)\circ\widehat{\mathbf{V}}\left(n\right)\right)\right)^{2}\right]<c_{1},\label{eq:bounded_c}
\end{equation}
as stated in $\left(b\right)$. We can say that \eqref{eq:e_1-1}
is true, as \eqref{eq:st_f1} implies that the probability that $\left|\sum_{n=1}^{N}\gamma_{n}e_{\boldsymbol{\Delta}}\left(n\right)\right|\geq\rho$
decreases with $\rho$.

In a similar way, \eqref{eq:e_1-1-1} can be proved. Since the additive
noise $\mathbf{Z}\left(n\right)$ is i.i.d., $\sum_{n=1}^{N}\gamma_{n}e_{\mathbf{Z}}^{\left(\mathrm{I}\right)}\left(n\right)$
is also martingale. Use the same step as in \eqref{eq:st_f1}, we
mainly need to verify whether $\mathbb{E}_{\mathbf{Z}}[(e_{\mathbf{Z}}^{\left(\mathrm{I}\right)}\left(n\right))^{2}]$
is bounded, which is straightforward to prove by the assumption that
$\mathbf{Z}\left(n\right)$ has bounded variance.

\subsection{\label{subsec:Proof-rate-i}Proof of Theorem \ref{prop:rate_incomp}}

By definition $D_{n}^{\left(\textrm{I}\right)}=\mathbb{E}\left[d_{n}^{\left(\textrm{I}\right)}\right]$,
we perform the expectation of \eqref{eq:dn+1_n_mxli} to get 
\begin{align}
D_{n+1}^{\left(\textrm{I}\right)} & \leq D_{n}^{\left(\textrm{I}\right)}+\gamma_{n}\mathbb{E}\left[\textrm{tr}\left(\left(\mathbf{X}\left(n\right)-\mathbf{X}^{*}\right)\boldsymbol{\Delta}\left(n\right)\circ\widehat{\mathbf{V}}\left(n\right)\right)\right]\nonumber \\
 & \qquad\qquad+\gamma_{n}^{2}\mathbb{E}\left[\left\Vert \boldsymbol{\Delta}\left(n\right)\circ\widehat{\mathbf{V}}\left(n\right)\right\Vert _{\infty}^{2}\right].\label{eq:dn+1_n}
\end{align}

Since the elements of $\boldsymbol{\Delta}_{k}\left(n\right)$ follow
Bernoulli distribution, it is easy to evaluate 
\begin{align}
 & \mathbb{E}\left[\textrm{tr}\left(\left(\mathbf{X}\left(n\right)-\mathbf{X}^{*}\right)\boldsymbol{\Delta}\left(n\right)\circ\widehat{\mathbf{V}}\left(n\right)\right)\right]\nonumber \\
 & =p\mathbb{E}\left[\textrm{tr}\left(\left(\mathbf{X}\left(n\right)-\mathbf{X}^{*}\right)\left(\mathbf{V}\left(n\right)+\mathbf{Z}\left(n\right)\right)\right)\right]\nonumber \\
 & \overset{\left(a\right)}{=}p\mathbb{E}\left[\textrm{tr}\left(\left(\mathbf{X}\left(n\right)-\mathbf{X}^{*}\right)\mathbf{V}\left(n\right)\right)\right]\nonumber \\
 & \overset{\left(b\right)}{\leq}-p_{\textrm{I}}BD_{n}^{\left(\textrm{I}\right)},\label{eq:st1}
\end{align}
in which $\left(a\right)$ is due to the fact that $\mathbf{Z}\left(n\right)$
has zero-mean elements and $\left(b\right)$ is by \eqref{eq:strong_stable}. 

Meanwhile, we evaluate
\begin{align}
\mathbb{E}\left[\left\Vert \widehat{\mathbf{V}}\left(n\right)\right\Vert _{\infty}^{2}\right] & \leq\mathbb{E}\left[\sum_{m}\left|\textrm{eig}_{m}\left(\widehat{\mathbf{V}}\left(n\right)\right)\right|^{2}\right]\nonumber \\
 & =\mathbb{E}\left[\sum_{i,j}\left|\widehat{V}_{i,j}\left(n\right)\right|^{2}\right]\leq C,\label{eq:st_0}
\end{align}
which comes from the bound in \eqref{eq:bound_V}. Similar to \eqref{eq:st_0},
we have 
\begin{align}
\mathbb{E}\left[\left\Vert \boldsymbol{\Delta}\left(n\right)\circ\widehat{\mathbf{V}}\left(n\right)\right\Vert _{\infty}^{2}\right] & \leq\mathbb{E}\left[\sum_{i,j}\left|\delta_{i,j}\left(n\right)\widehat{V}_{i,j}\left(n\right)\right|^{2}\right]\nonumber \\
 & =p_{\textrm{I}}\mathbb{E}\left[\sum_{i,j}\left|\widehat{V}_{i,j}\left(n\right)\right|^{2}\right]\nonumber \\
 & \leq p_{\textrm{I}}C.\label{eq:st_2}
\end{align}

Combining \eqref{eq:dn+1_n}, \eqref{eq:st1}, and \eqref{eq:st_2},
we obtain
\begin{equation}
D_{n+1}^{\left(\textrm{I}\right)}\leq\left(1-p_{\textrm{I}}B\gamma_{n}\right)D_{n}^{\left(\textrm{I}\right)}+p_{\textrm{I}}C\gamma_{n}^{2}.\label{eq:dd_n_n+1_I}
\end{equation}
Our aim is to show the existence of a bounded constant $\lambda$
such that $D_{n}^{\left(\textrm{I}\right)}\leq\lambda\gamma_{n}$.
The proof is by induction. 

Obviously, we have $D_{1}^{\left(\textrm{I}\right)}\leq\lambda\gamma_{1}$
if $\lambda\geq D_{1}^{\left(\textrm{I}\right)}/\gamma_{1}$. Then
under the condition that $D_{n}^{\left(\textrm{I}\right)}\leq\lambda\gamma_{n}$,
we need to verify whether $D_{n+1}^{\left(\textrm{I}\right)}\leq\lambda\gamma_{n+1}$.
As $\gamma_{n}\leq\gamma_{1}<\frac{1}{p_{\textrm{I}}B}$ $\forall n$,
from \eqref{eq:dd_n_n+1_I}, we need to have
\begin{align}
D_{n+1}^{\left(\textrm{I}\right)} & \leq\left(1-p_{\textrm{I}}B\gamma_{n}\right)\lambda\gamma_{n}+p_{\textrm{I}}C\gamma_{n}^{2}\nonumber \\
 & \leq\lambda\gamma_{n+1},\label{eq:st_3}
\end{align}
meaning that $\lambda$ should be such that
\begin{equation}
\lambda\geq\frac{p_{\textrm{I}}C}{p_{\textrm{I}}B-\max_{n}\left(\frac{\gamma_{n}-\gamma_{n+1}}{\gamma_{n}^{2}}\right)}=\frac{p_{\textrm{I}}C}{p_{\textrm{I}}B-\varepsilon}
\end{equation}
under the condition $p_{\textrm{I}}B>\varepsilon$. In this way, we
conclude that $D_{n}^{\left(\textrm{I}\right)}\leq\lambda\gamma_{n}$
if $\lambda\geq\max\left\{ \frac{D_{1}^{\left(\textrm{I}\right)}}{\gamma_{1}},\frac{p_{\textrm{I}}C}{p_{\textrm{I}}B-\varepsilon}\right\} $,
which concludes the proof.

\subsection{\label{subsec:Proof-of-Lemma_boundi}Proof of Lemma \ref{lem:bound_i}}

Consider the first situation where $\nu=1$, \eqref{eq:gx_bound}
holds as 
\begin{equation}
g(x)=x^{-1}\left(1-\frac{1}{1+x}\right)=\frac{1}{1+x}\leq1.
\end{equation}

In the other situation, \emph{i.e.}, $\nu\in\left(0.5,1\right),$
we evaluate the derivative of $g\left(x\right)$, \emph{i.e.}, $g'\left(x\right)=\nu x^{-\nu-1}h\left(x\right)$
with 
\begin{equation}
h\left(x\right)=\left(2x+1\right)\left(1+x\right)^{-\nu-1}-1.
\end{equation}
Thus the monotonicity of $g\left(x\right)$ depends on whether $h(x)$
is positive or negative. We further evaluate 
\begin{align}
h'(x) & =2\nu\left(1+x\right)^{-\nu-2}\left(\frac{1-\nu}{2\nu}-x\right),
\end{align}
from which we deduce that $h(x)$ is an increasing function as $x\in\left(0,\frac{1-\nu}{2\nu}\right]$
and $h(x)$ decreases while $x\in\left[\frac{1-\nu}{2\nu},1\right]$.
Note that $\frac{1-\nu}{2\nu}\in\left(0,1\right)$. It is easy to
get $\lim_{x\rightarrow0}h\left(x\right)=0$ and $h(\frac{1-\nu}{2\nu})>0$.
Besides, $h(1)=3\cdot2^{-\nu-1}-1$ depends on the value of $\nu$.

If $\nu\in(0.5,\log_{2}1.5]$, then $h\left(1\right)\geq0$, which
implies that $g\left(x\right)$ is an increasing function over $(0,1]$,
thus $g\left(x\right)\leq g\left(1\right)=1-2^{-\nu}$.

If $\nu\in(\log_{2}1.5,1)$, then $h\left(1\right)<0$. Based on the
monotonicity of $h(x)$, we concludes that there exists a single point
$x_{0}\in\left(\frac{1-\nu}{2\nu},1\right)$ such that $h(x_{0})=0$
and $g(x)\leq g\left(x_{0}\right)$. From $h(x_{0})=0$, we get $\left(1+x_{0}\right)^{-\nu}=\frac{1+x_{0}}{1+2x_{0}}$,
thus 
\begin{align}
g\left(x\right) & \leq g\left(x_{0}\right)=x_{0}^{-\nu}\left(1-\frac{1+x_{0}}{1+2x_{0}}\right)\nonumber \\
 & =\frac{x_{0}^{1-\nu}}{1+2x_{0}}\leq\nu\left(\frac{1-\nu}{2\nu}\right)^{1-\nu},
\end{align}
which takes the equality as $x_{0}=\frac{1-\nu}{2\nu}$.

\subsection{\label{subsec:Proof-of-sum}Proof of Lemma \ref{lem:gamma_sum}}

For any $k$ and $n$, we evaluate 
\begin{align}
 & \mathbb{E}\left[\gamma_{n_{k}}\eta_{k}\left(n\right)\right]\nonumber \\
 & =\mathbb{P}\left[\eta_{k}\left(n\right)=1\right]\mathbb{E}\left[\gamma_{\widetilde{n}_{k}+\eta_{k}\left(n\right)}\eta_{k}\left(n\right)\mid\eta_{k}\left(n\right)=1\right]\nonumber \\
 & =p_{\textrm{S}}\sum_{\ell=1}^{n}\gamma_{\ell}\mathbb{P}\left[\widetilde{n}_{k}=\ell-1\right]\nonumber \\
 & =\sum_{\ell=1}^{n}\gamma_{\ell}p_{\textrm{S}}^{\ell}\left(1-p_{\textrm{S}}\right)^{n-\ell}\binom{n-1}{\ell-1},\label{eq:aver1}
\end{align}
recall that $\eta_{k}\left(n\right)\sim\mathcal{B}\left(1,p_{\textrm{S}}\right)$
and $\widetilde{n}_{k}\sim\mathcal{B}\left(n-1,p_{\textrm{S}}\right)$.
We find that $\overline{\gamma}_{n}$ defined in \eqref{eq:ga_1}
represents $\mathbb{E}\left[\gamma_{n_{k}}\eta_{k}\left(n\right)\right]$.
Similarly, we have 
\begin{align}
 & \mathbb{E}\left[\left(\gamma_{n_{k}}\eta_{k}\left(n\right)\right)^{2}\right]\nonumber \\
 & =\mathbb{P}\left[\eta_{k}\left(n\right)=1\right]\mathbb{E}\left[\left.\left(\gamma_{\widetilde{n}_{k}+\eta_{k}\left(n\right)}\eta_{k}\left(n\right)\right)^{2}\right|\eta_{k}\left(n\right)=1\right]\nonumber \\
 & =p_{\textrm{S}}\sum_{\ell=1}^{n}\gamma_{\ell}^{2}\mathbb{P}\left[\widetilde{n}_{k}=\ell-1\right]\nonumber \\
 & =\sum_{\ell=1}^{n}\gamma_{\ell}^{2}p_{\textrm{S}}^{\ell}\left(1-p_{\textrm{S}}\right)^{n-\ell}\binom{n-1}{\ell-1},\label{eq:aver2}
\end{align}
then $\mathring{\gamma}_{n}=\sqrt{\mathbb{E}\left[\left(\gamma_{n_{k}}\eta_{k}\left(n\right)\right)^{2}\right]}$
are also obtained.

In order to prove \eqref{eq:gamma_aver_sum} and \eqref{eq:gamma_aversquar},
we first present a useful lemma in what follows with its proof presented
in the end of this appendix.
\begin{lem}
\label{lem:sum_equi}Consider an arbitrary sequence $\left\{ a_{n}\right\} $
and define 
\begin{align}
\overline{a}_{n} & =\sum_{\ell=1}^{n}a_{\ell}\left(1-p\right)^{\ell}p^{n-\ell}\binom{n-1}{\ell-1},
\end{align}
with $p\in\left[0,1\right]$, we always have 
\begin{equation}
\sum_{n=1}^{\infty}\overline{a}_{n}=\sum_{n=1}^{\infty}a_{n}.
\end{equation}
\end{lem}
Replace $a_{n}$ by $\gamma_{n}$ and $p$ by $1-p_{\textrm{S}}$,
we get that $\sum_{n=1}^{\infty}\overline{\gamma}_{n}=\sum_{n=1}^{\infty}\gamma_{n}$,
then \eqref{eq:gamma_aver_sum} can be proved as $\sum_{n=1}^{\infty}\gamma_{n}\rightarrow\infty$.

Replace $a_{n}$ by $\gamma_{n}^{2}$ and $p$ by $1-p_{\textrm{S}}$,
we have $\sum_{n=1}^{\infty}\mathring{\gamma}_{n}^{2}=\sum_{n=1}^{\infty}\gamma_{n}^{2}$.
Due to the fact that 
\begin{equation}
\mathring{\gamma}_{n}^{2}=\mathbb{E}\left[\left(\gamma_{n_{k}}\eta_{k}\left(n\right)\right)^{2}\right]\geq\left(\mathbb{E}\left[\gamma_{n_{k}}\eta_{k}\left(n\right)\right]\right)^{2}=\overline{\gamma}_{n}^{2},
\end{equation}
we can finally justify \eqref{eq:gamma_aversquar}, with the assumption
that $\sum_{n=1}^{\infty}\gamma_{n}^{2}<\infty$.

The proof of Lemma~\ref{lem:sum_equi} is presented in the following.
\begin{IEEEproof}
We evaluate
\begin{align}
\sum_{n=1}^{\infty}\overline{a}_{n} & =\sum_{n=0}^{\infty}\overline{a}_{n+1}\nonumber \\
 & =\left(1-p\right)\sum_{n=0}^{\infty}\sum_{\ell=0}^{n}a_{\ell+1}\left(1-p\right)^{\ell}p^{n-\ell}\binom{n}{\ell}\nonumber \\
 & \overset{\left(a\right)}{=}\left(1-p\right)\sum_{\ell=0}^{\infty}a_{\ell+1}\sum_{n=\ell}^{\infty}\left(1-p\right)^{\ell}p^{n-\ell}\binom{n}{\ell}\nonumber \\
 & \overset{\left(b\right)}{=}\left(1-p\right)\sum_{\ell=0}^{\infty}a_{\ell+1}w_{\ell},\label{eq:a_aver_s}
\end{align}
where $\left(a\right)$ is obtained by changing the order of summation
and $\left(b\right)$ is by introducing $w_{\ell}=\sum_{n=\ell}^{\infty}\left(1-p\right)^{\ell}p^{n-\ell}\binom{n}{\ell}$,
which is independent of $n$. We should then focus on the expression
of $w_{\ell}$. We have 
\begin{align}
w_{\ell} & \overset{\left(a\right)}{=}\sum_{m=0}^{\infty}\left(1-p\right)^{\ell}p^{m}\binom{m+\ell}{\ell}\nonumber \\
 & =\frac{\left(1-p\right)^{\ell}}{\ell!}\sum_{m=0}^{\infty}\left(m+\ell\right)\left(m+\ell-1\right)\cdots\left(m+1\right)p^{m}\nonumber \\
 & \overset{\left(b\right)}{=}\frac{\left(1-p\right)^{\ell}}{\ell!}\sum_{m=0}^{\infty}\left(p^{m+\ell}\right)^{\left(\ell\right)}=\frac{\left(1-p\right)^{\ell}}{\ell!}\left(\sum_{m=0}^{\infty}p^{m+\ell}\right)^{\left(\ell\right)}\nonumber \\
 & =\frac{\left(1-p\right)^{\ell}}{\ell!}\left(p^{\ell}\left(1-p\right)^{-1}\right)^{\left(\ell\right)},\label{eq:wl1}
\end{align}
in which $\left(a\right)$ is by change of variable $m=n-\ell$; in
$\left(b\right)$, we denote $\left(p^{m+\ell}\right)^{\left(\ell\right)}$
as the $\ell$-th order derivative of the function $p^{m+\ell}$. 

By applying the general Leibniz rule to evaluate the $\ell$-th order
derivative of the function $p^{\ell}\left(1-p\right)^{-1}$, \eqref{eq:wl1}
can be written as
\begin{align}
w_{\ell} & =\frac{\left(1-p\right)^{\ell}}{\ell!}\sum_{k=0}^{\ell}\binom{\ell}{k}\left(\left(1-p\right)^{-1}\right)^{\left(k\right)}\left(p^{\ell}\right)^{\left(\ell-k\right)}\nonumber \\
 & \overset{\left(a\right)}{=}\frac{\left(1-p\right)^{\ell}}{\ell!}\sum_{k=0}^{\ell}\binom{\ell}{k}k!\left(1-p\right)^{-k-1}\frac{\ell!}{k!}p^{k}\nonumber \\
 & =\sum_{k=0}^{\ell}\binom{\ell}{k}\left(1-p\right)^{\ell-k-1}p^{k}\nonumber \\
 & =\left(1-p\right)^{-1},\label{eq:wl2}
\end{align}
where $\left(a\right)$ is obtained by $\left(\left(1-p\right)^{-1}\right)^{\left(k\right)}=k!\left(1-p\right)^{-k-1}$
and $\left(p^{\ell}\right)^{\left(\ell-k\right)}=\frac{\ell!}{k!}p^{k}$.
We deduce that $w_{\ell}$ is in fact a constant for any $\ell$.

Combine \eqref{eq:a_aver_s} and \eqref{eq:wl2}, we get that
\begin{align}
\sum_{n=1}^{\infty}\overline{a}_{n} & =\left(1-p\right)\sum_{\ell=0}^{\infty}a_{\ell+1}\left(1-p\right)^{-1}\nonumber \\
 & =\sum_{\ell=0}^{\infty}a_{\ell+1},
\end{align}
which concludes the proof.
\end{IEEEproof}

\subsection{\label{subsec:Proof-of-conv_mxl_s}Proof of Theorem~\ref{thm:mxls_conv}}

Introduce a all-ones matrix $\mathbf{1}_{k}$ of the same shape as
$\mathbf{V}_{k}$ and denote $\boldsymbol{\Gamma}(n)=\textrm{diag}\left(\gamma_{n_{k}}\eta_{k}\left(n\right)\mathbf{1}_{k}\right)_{k=1}^{K}$.
From \eqref{eq:mxl_s}, we have $\mathbf{Y}^{\left(\textrm{S}\right)}\left(n\right)=\mathbf{Y}^{\left(\textrm{S}\right)}\left(n-1\right)+\boldsymbol{\Gamma}(n)\circ\widehat{\mathbf{V}}\left(n\right)$. 

Similar to \eqref{eq:dn+1_n_mxli}, we have
\begin{align}
d_{n+1}^{\left(\textrm{S}\right)} & \leq d_{n}^{\left(\textrm{S}\right)}+\textrm{tr}\left(\left(\mathbf{X}^{\left(\textrm{S}\right)}\left(n\right)-\mathbf{X}^{*}\right)\boldsymbol{\Gamma}(n)\circ\widehat{\mathbf{V}}\left(n\right)\right)\nonumber \\
 & \qquad\qquad+\left\Vert \boldsymbol{\Gamma}(n)\circ\widehat{\mathbf{V}}\left(n\right)\right\Vert _{\infty}^{2}.\label{eq:dn+1_n-1-1}
\end{align}
As $e_{\boldsymbol{\Delta}}\left(n\right)$ and $e_{\mathbf{Z}}^{\left(\mathrm{I}\right)}\left(n\right)$
defined in Appendix~\ref{subsec:Proof-convmxli}, we define

\begin{align}
e_{\boldsymbol{\Gamma}}\left(n\right) & =\textrm{tr}\left(\left(\mathbf{X}^{\left(\textrm{S}\right)}\left(n\right)-\mathbf{X}^{*}\right)\boldsymbol{\Gamma}(n)\circ\widehat{\mathbf{V}}\left(n\right)\right)\nonumber \\
 & \qquad-\mathbb{E}_{\boldsymbol{\Gamma}}\left[\textrm{tr}\left(\left(\mathbf{X}^{\left(\textrm{S}\right)}\left(n\right)-\mathbf{X}^{*}\right)\boldsymbol{\Gamma}(n)\circ\widehat{\mathbf{V}}\left(n\right)\right)\right]\nonumber \\
 & =\textrm{tr}\left(\left(\mathbf{X}^{\left(\textrm{S}\right)}\left(n\right)-\mathbf{X}^{*}\right)\boldsymbol{\Gamma}(n)\circ\widehat{\mathbf{V}}\left(n\right)\right)\nonumber \\
 & \qquad-\overline{\gamma}_{n}\textrm{tr}\left(\left(\mathbf{X}^{\left(\textrm{S}\right)}\left(n\right)-\mathbf{X}^{*}\right)\widehat{\mathbf{V}}\left(n\right)\right),\label{eq:e_tau}
\end{align}
and 
\begin{equation}
e_{\mathbf{Z}}^{\left(\mathrm{S}\right)}\left(n\right)=\overline{\gamma}_{n}\textrm{tr}\left(\left(\mathbf{X}^{\left(\textrm{S}\right)}\left(n\right)-\mathbf{X}^{*}\right)\mathbf{Z}\left(n\right)\right),\label{eq:e_zs}
\end{equation}
in which $\overline{\gamma}_{n}=\mathbb{E}\left[\gamma_{n_{k}}\eta_{k}\left(n\right)\right]$
as defined in \eqref{eq:ga_1}. By introducing \eqref{eq:e_tau}-\eqref{eq:e_zs}
into \eqref{eq:dn+1_n-1-1}, we get 
\begin{align}
d_{n+1}^{\left(\textrm{S}\right)} & \leq d_{n}^{\left(\textrm{S}\right)}+\overline{\gamma}_{n}\textrm{tr}\left(\left(\mathbf{X}^{\left(\textrm{S}\right)}\left(n\right)-\mathbf{X}^{*}\right)\mathbf{V}\left(n\right)\right)\nonumber \\
 & +e_{\boldsymbol{\Gamma}}\left(n\right)+e_{\mathbf{Z}}^{\left(\mathrm{S}\right)}\left(n\right)+\left\Vert \boldsymbol{\Gamma}(n)\circ\widehat{\mathbf{V}}\left(n\right)\right\Vert _{\infty}^{2},\label{eq:dn+1_n_mxls-1-1}
\end{align}
from which we deduce that 
\begin{align}
d_{N+1}^{\left(\textrm{S}\right)} & \leq d_{1}^{\left(\textrm{S}\right)}+\sum_{n=1}^{N}\overline{\gamma}_{n}\textrm{tr}\left(\left(\mathbf{X}^{\left(\textrm{S}\right)}\left(n\right)-\mathbf{X}^{*}\right)\mathbf{V}\left(n\right)\right)+E_{N}^{\left(\mathrm{S}\right)},\label{eq:dn+1_n_mxls-1}
\end{align}
with 
\begin{align}
E_{N}^{\left(\mathrm{S}\right)} & =\sum_{n=1}^{N}\left(e_{\boldsymbol{\Gamma}}\left(n\right)+e_{\mathbf{Z}}^{\left(\mathrm{S}\right)}\left(n\right)+\left\Vert \boldsymbol{\Gamma}(n)\circ\widehat{\mathbf{V}}\left(n\right)\right\Vert _{\infty}^{2}\right).\label{eq:esum_s}
\end{align}
Similar to Lemma~\ref{lem:bound_E}, it is important to investigate
the property of $E_{N}^{\left(\mathrm{S}\right)}$. 
\begin{lem}
\label{lem:bound_E-s}As long as Assumptions A1-A2 hold, we have $\left|E_{N}^{\left(\mathrm{S}\right)}\right|<\infty$
a.s..
\end{lem}
\begin{IEEEproof}
See Appendix~\ref{subsec:Proof-of-e2}.
\end{IEEEproof}
With the result presented in Lemma~\ref{lem:bound_E-s}, as well
as the important property stated in Lemma~\ref{lem:gamma_sum}, \emph{i.e.},
$\sum_{n=1}^{\infty}\overline{\gamma}_{n}\rightarrow\infty$, the
rest part of the proof is the same as that in Appendix~\ref{subsec:Proof-convmxli},
since \eqref{eq:dn+1_n_mxls-1} has the same form compared with \eqref{eq:dn+1_n_mxli-2}. 

\subsection{\label{subsec:Proof-of-e2}Proof of Lemma~\ref{lem:bound_E-s}}

The proof steps presented in this appendix is similar to that in Appendix~\ref{subsec:Proof_sto_e1}.
We need to prove
\begin{align}
\lim_{N\rightarrow\infty}\sum_{n=1}^{N}\left\Vert \boldsymbol{\Gamma}(n)\circ\widehat{\mathbf{V}}\left(n\right)\right\Vert _{\infty}^{2} & <\infty,\label{eq:e_1-2}\\
\lim_{N\rightarrow\infty}\left|\sum_{n=1}^{N}e_{\mathbf{Z}}^{\left(\mathrm{S}\right)}\left(n\right)\right| & <\infty,\quad\mathrm{a.s.}\label{eq:e_1-1-1-1}\\
\lim_{N\rightarrow\infty}\left|\sum_{n=1}^{N}e_{\boldsymbol{\Gamma}}\left(n\right)\right| & <\infty,\quad\mathrm{a.s.}\label{eq:e_tau-1}
\end{align}

The proof of \eqref{eq:e_1-2} is quite similar to that of \eqref{eq:e_1}.
We focus on the demonstration of \eqref{eq:e_1-1-1-1} and \eqref{eq:e_tau-1}.

\subsubsection{Proof of \eqref{eq:e_1-1-1-1}}

Similar to $\sum_{n=1}^{N}e_{\mathbf{Z}}^{\left(\mathrm{I}\right)}\left(N\right)$,
$\sum_{n=1}^{N}e_{\mathbf{Z}}^{\left(\mathrm{S}\right)}\left(N\right)$
is also martingale. We apply Doob's inequality again to have, for
any $\rho>0$,
\begin{align}
 & \mathbb{P}\left[\sup_{N}\left|\sum_{n=1}^{N}e_{\mathbf{Z}}^{\left(\mathrm{S}\right)}\left(n\right)\right|\geq\rho\right]\nonumber \\
 & \leq\rho^{-2}\mathbb{E}_{\mathbf{Z}}\left[\left(\sum_{n=1}^{N}e_{\mathbf{Z}}^{\left(\mathrm{S}\right)}\left(n\right)\right)^{2}\right]=\rho^{-2}\sum_{n=1}^{N}\mathbb{E}_{\mathbf{Z}}\left[\left(e_{\mathbf{Z}}^{\left(\mathrm{S}\right)}\left(n\right)\right)^{2}\right]\nonumber \\
 & =\rho^{-2}\sum_{n=1}^{N}\overline{\gamma}_{n}^{2}\mathbb{E}_{\mathbf{Z}}\left[\left(\textrm{tr}\left(\left(\mathbf{X}^{\left(\textrm{S}\right)}\left(n\right)-\mathbf{X}^{*}\right)\mathbf{Z}\left(n\right)\right)\right)^{2}\right]\nonumber \\
 & \overset{}{\leq}\rho^{-2}c_{2}\sum_{n=1}^{N}\overline{\gamma}_{n}^{2},
\end{align}
where we have, similar to \eqref{eq:bounded_c},
\begin{equation}
\mathbb{E}_{\mathbf{Z}}\left[\left(\textrm{tr}\left(\left(\mathbf{X}^{\left(\mathrm{S}\right)}\left(n\right)-\mathbf{X}_{k}^{*}\right)\mathbf{Z}_{k}\left(n\right)\right)\right)^{2}\right]\leq c_{2}<\infty.\label{eq:bounded_c2}
\end{equation}
According to Lemma~\ref{lem:gamma_sum}, $\sum_{n=1}^{\infty}\overline{\gamma}_{n}^{2}$
is also bounded, therefore we conclude that $\left|\sum_{n=1}^{N}e_{\mathbf{Z}}^{\left(\mathrm{S}\right)}\left(n\right)\right|<\infty$
almost surely as $N\rightarrow\infty$.

\subsubsection{Proof of \eqref{eq:e_tau-1}}

Due to the fact that the step-size of each user is generated independently
and randomly, we can easily show that $\sum_{n=1}^{N}e_{\boldsymbol{\Gamma}}\left(n\right)$
is also martingale. We have, by Doob's inequality, for any $\rho>0$,
\begin{align}
 & \mathbb{P}\left[\sup_{N}\left|\sum_{n=1}^{N}e_{\boldsymbol{\Gamma}}\left(n\right)\right|\geq\rho\right]\nonumber \\
 & \leq\rho^{-2}\mathbb{E}_{\boldsymbol{\Gamma}}\left[\left(\sum_{n=1}^{N}e_{\boldsymbol{\Gamma}}\left(n\right)\right)^{2}\right]=\rho^{-2}\sum_{n=1}^{N}\mathbb{E}_{\boldsymbol{\Gamma}}\left[e_{\boldsymbol{\Gamma}}^{2}\left(n\right)\right]\nonumber \\
 & \leq\rho^{-2}\sum_{n=1}^{N}\mathbb{E}_{\boldsymbol{\Gamma}}\left[\left(\textrm{tr}\left(\left(\mathbf{X}^{\left(\textrm{S}\right)}\left(n\right)-\mathbf{X}^{*}\right)\boldsymbol{\Gamma}(n)\circ\widehat{\mathbf{V}}\left(n\right)\right)\right)^{2}\right]\nonumber \\
 & \overset{\left(a\right)}{=}\rho^{-2}\sum_{n=1}^{N}\mathbb{E}_{\boldsymbol{\Gamma}}\left[\left(\sum_{k=1}^{K}\gamma_{n_{k}}\eta_{k}\left(n\right)\right.\right.\nonumber \\
 & \qquad\qquad\qquad\qquad\left.\left.\cdot\textrm{tr}\left(\left(\mathbf{X}_{k}^{\left(\textrm{S}\right)}\left(n\right)-\mathbf{X}_{k}^{*}\right)\widehat{\mathbf{V}}_{k}\left(n\right)\right)\right)^{2}\right]\nonumber \\
 & \overset{\left(b\right)}{=}\rho^{-2}\sum_{n=1}^{N}\sum_{k=1}^{K}\mathbb{E}_{\boldsymbol{\Gamma}}\left[\left(\gamma_{n_{k}}\eta_{k}\left(n\right)\right)^{2}\right]\nonumber \\
 & \qquad\qquad\qquad\qquad\cdot\left(\textrm{tr}\left(\left(\mathbf{X}_{k}^{\left(\textrm{S}\right)}\left(n\right)-\mathbf{X}_{k}^{*}\right)\widehat{\mathbf{V}}_{k}\left(n\right)\right)\right)^{2}\nonumber \\
 & =\rho^{-2}\sum_{n=1}^{N}\mathring{\gamma}_{n}^{2}\sum_{k=1}^{K}\left(\textrm{tr}\left(\left(\mathbf{X}_{k}^{\left(\textrm{S}\right)}\left(n\right)-\mathbf{X}_{k}^{*}\right)\widehat{\mathbf{V}}_{k}\left(n\right)\right)\right)^{2}\nonumber \\
 & \overset{\left(c\right)}{\leq}\rho^{-2}c_{3}\sum_{n=1}^{N}\mathring{\gamma}_{n}^{2},
\end{align}
where $\left(a\right)$ is comes from the fact that $\mathbf{X}$,
$\boldsymbol{\Gamma}$, and $\widehat{\mathbf{V}}$ are all block-diagonal
matrices with the same shape; $\left(b\right)$ is by $\mathbb{E}_{\boldsymbol{\Gamma}}\left[\eta_{k_{1}}\left(n\right)\eta_{k_{2}}\left(n\right)\right]=0$
for any $k_{1}\neq k_{2}$; in $(c)$, there exists $c_{3}<\infty$
such that
\begin{equation}
\sum_{k=1}^{K}\left(\textrm{tr}\left(\left(\mathbf{X}_{k}^{\left(\textrm{S}\right)}\left(n\right)-\mathbf{X}_{k}^{*}\right)\widehat{\mathbf{V}}_{k}\left(n\right)\right)\right)^{2}\leq c_{3},
\end{equation}
as the entries of $\widehat{\mathbf{V}}$ and $\mathbf{X}$ have bounded
values. In the end, \eqref{eq:e_tau-1} can be justified as $\sum_{n=1}^{N}\mathring{\gamma}_{n}^{2}<\infty$
by Lemma~\ref{lem:gamma_sum}.

\subsection{\label{subsec:Proof-2}Proof of Theorem \ref{prop:rate_asy}}

Perform the expectation of \eqref{eq:dn+1_n-1-1}, we have 

\begin{align}
D_{n+1}^{\left(\textrm{S}\right)} & \leq D_{n}^{\left(\textrm{S}\right)}+\mathbb{E}\left[\textrm{tr}\left(\left(\mathbf{X}^{\left(\textrm{S}\right)}\left(n\right)-\mathbf{X}^{*}\right)\boldsymbol{\Gamma}(n)\circ\widehat{\mathbf{V}}\left(n\right)\right)\right]\nonumber \\
 & \qquad\qquad+\mathbb{E}\left[\left\Vert \boldsymbol{\Gamma}(n)\circ\widehat{\mathbf{V}}\left(n\right)\right\Vert _{\infty}^{2}\right].\label{eq:dn+1_n-1}
\end{align}
Recall that $\overline{\gamma}_{n}$ defined in \eqref{eq:ga_1} represents
$\mathbb{E}\left[\gamma_{n_{k}}\eta_{k}\left(n\right)\right]$. 
\begin{align}
 & \mathbb{E}\left[\textrm{tr}\left(\left(\mathbf{X}^{\left(\textrm{S}\right)}\left(n\right)-\mathbf{X}^{*}\right)\boldsymbol{\Gamma}(n)\circ\widehat{\mathbf{V}}\left(n\right)\right)\right]\nonumber \\
 & =\mathbb{E}\left[\textrm{tr}\left(\left(\mathbf{X}^{\left(\textrm{S}\right)}\left(n\right)-\mathbf{X}^{*}\right)\textrm{diag}\left(\gamma_{n_{k}}\eta_{k}\left(n\right)\widehat{\mathbf{V}}_{k}\left(n\right)\right)_{k=1}^{K}\right)\right]\nonumber \\
 & =\overline{\gamma}_{n}\mathbb{E}\left[\textrm{tr}\left(\left(\left(\mathbf{X}^{\left(\textrm{S}\right)}\mathbf{X}\right)\left(n\right)-\mathbf{X}^{*}\right)\widehat{\mathbf{V}}\left(n\right)\right)\right]\nonumber \\
 & \leq-B\overline{\gamma}_{n}D_{n}^{\left(\textrm{S}\right)},
\end{align}
and
\begin{align}
\mathbb{E}\left[\left\Vert \boldsymbol{\Gamma}(n)\circ\widehat{\mathbf{V}}\left(n\right)\right\Vert _{\infty}^{2}\right] & \leq\mathbb{E}\left[\sum_{i,j,k}\left(\gamma_{n_{k}}\eta_{k}\left(n\right)\right)^{2}\left|\widehat{V}_{k,i,j}\left(n\right)\right|^{2}\right]\nonumber \\
 & =\mathring{\gamma}_{n}^{2}\mathbb{E}\left[\sum_{i,j}\left|\widehat{V}_{i,j}\left(n\right)\right|^{2}\right]\leq C\mathring{\gamma}_{n}^{2}.
\end{align}
Thus \eqref{eq:dn+1_n-1} leads to
\begin{equation}
D_{n+1}^{\left(\textrm{S}\right)}\leq\left(1-B\overline{\gamma}_{n}\right)D_{n}^{\left(\textrm{S}\right)}+C\mathring{\gamma}_{n}^{2}.
\end{equation}
The rest of the proof is still by induction, the steps are similar
to that in Section \ref{subsec:Proof-2}. We mainly need to find $\nu$
such that $D_{n}^{\left(\textrm{S}\right)}\leq\mu\mathring{\gamma}_{n}^{2}/\overline{\gamma}_{n}$.
Since $B\overline{\gamma}_{n}\leq B\gamma_{1}\leq1$, we should have
\begin{equation}
D_{n+1}^{\left(\textrm{S}\right)}\leq\left(1-B\overline{\gamma}_{n}\right)\mu\frac{\mathring{\gamma}_{n}^{2}}{\overline{\gamma}_{n}}+C\mathring{\gamma}_{n}^{2}\leq\mu\frac{\mathring{\gamma}_{n+1}^{2}}{\overline{\gamma}_{n+1}},
\end{equation}
leading to
\begin{equation}
\mu\geq\frac{C}{B-\max_{n}\left(\frac{1}{\mathring{\gamma}_{n}^{2}}\left(\frac{\mathring{\gamma}_{n}^{2}}{\overline{\gamma}_{n}}-\frac{\mathring{\gamma}_{n+1}^{2}}{\overline{\gamma}_{n+1}}\right)\right)}=\frac{C}{B-\epsilon}.
\end{equation}
under the condition that $B>\epsilon$, which concludes the proof.

\subsection{\label{subsec:Proof-of-Lemma_bound}Proof of Lemma \ref{lem:rate_bound}}

For any $n=1,2,\ldots$, consider a random sequence $W_{n}$ with
$W_{n}\sim\mathcal{B}\left(n,p_{\textrm{S}}\right)$. By definition,
we have 
\[
\overline{\gamma}_{n+1}=p_{\textrm{S}}\mathbb{E}\left[\gamma_{W_{n}+1}\right],\quad\mathring{\gamma}_{n+1}^{2}=p_{\textrm{S}}\mathbb{E}\left[\gamma_{W_{n}+1}^{2}\right].
\]
Due to the fact that $\gamma_{\ell}$ is convex over $\ell$, we can
obtain a lower bound of $\overline{\gamma}_{n+1}$ using Jensen's
inequality, \emph{i.e.}, 
\begin{equation}
\overline{\gamma}_{n+1}\geq p_{\textrm{S}}\gamma_{\left\lfloor \mathbb{E}\left[W_{n}\right]\right\rfloor +1+1}=p_{\textrm{S}}\gamma_{\left\lfloor p_{\textrm{S}}n\right\rfloor +2}.\label{eq:lowerb}
\end{equation}
Now we need to find an upper bound of $\mathring{\gamma}_{n+1}^{2}$.
Consider an arbitrary $0<\xi<1$, we have 
\begin{align}
\mathring{\gamma}_{n+1}^{2} & =p_{\textrm{S}}\sum_{\ell=0}^{n}\mathbb{P}\left[W_{n}=\ell\right]\gamma_{\ell+1}^{2}\nonumber \\
 & \overset{\left(a\right)}{\leq}p_{\textrm{S}}\left(\sum_{\ell=0}^{\left\lfloor \left(1-\xi\right)p_{\textrm{S}}n\right\rfloor }\mathbb{P}\left[W_{n}=\ell\right]\gamma_{1}^{2}\right.\nonumber \\
 & \qquad+\left.\sum_{\ell=\left\lfloor \left(1-\xi\right)p_{\textrm{S}}n\right\rfloor +1}^{n}\mathbb{P}\left[W_{n}=\ell\right]\gamma_{\left\lfloor \left(1-\xi\right)p_{\textrm{S}}n\right\rfloor +2}^{2}\right)\nonumber \\
 & =p_{\textrm{S}}\left(\gamma_{1}^{2}\mathbb{P}\left[W_{n}\leq\left\lfloor \left(1-\xi\right)p_{\textrm{S}}n\right\rfloor \right]+\right.\nonumber \\
 & \qquad+\left.\gamma_{\left\lfloor \left(1-\xi\right)p_{\textrm{S}}n\right\rfloor +2}^{2}\mathbb{P}\left[W_{n}>\left\lfloor \left(1-\xi\right)p_{\textrm{S}}n\right\rfloor \right]\right)\nonumber \\
 & \overset{\left(b\right)}{<}p_{\textrm{S}}\left(\exp\left(-\frac{1}{2}\xi^{2}p_{\textrm{S}}n\right)\gamma_{1}^{2}+\gamma_{\left\lfloor \left(1-\xi\right)p_{\textrm{S}}n\right\rfloor +2}^{2}\right)\label{eq:upperb}
\end{align}
where $(a)$ is by the monotonicity of $\gamma_{\ell}$, \emph{i.e.},
$\gamma_{\ell+1}\leq1$ for any $\ell\geq0$ and $\gamma_{\ell+1}\leq\gamma_{\left\lfloor \left(1-\xi\right)np_{\textrm{S}}\right\rfloor +2}$
for any $\ell\geq\left\lfloor \left(1-\xi\right)p_{\textrm{S}}n\right\rfloor +1$;
in $\left(b\right)$, we consider $\mathbb{P}\left[W_{n}>\left\lfloor \left(1-\xi\right)p_{\textrm{S}}n\right\rfloor \right]<1$
and we apply Chernoff Bound, \emph{i.e.}, 
\begin{equation}
\mathbb{P}\left[W_{n}\leq\left(1-\xi\right)\mathbb{E}\left[W_{n}\right]\right]\leq\exp\left(-\frac{1}{2}\xi^{2}\mathbb{E}\left[W_{n}\right]\right),
\end{equation}
recall that $\mathbb{E}\left[W_{n}\right]=p_{\textrm{S}}n$. Combining
\eqref{eq:lowerb} and \eqref{eq:upperb}, we can finally obtain \eqref{eq:rate_a}.

\bibliographystyle{IEEEtran}
\bibliography{ref,BiblioWenjie}

\end{document}